%% file: neurips_2026.tex
\RequirePackage{fix-cm}
\documentclass[letterpaper,11pt]{article}
\PassOptionsToPackage{table}{xcolor}

\usepackage[utf8]{inputenc}
\usepackage[T1]{fontenc}
\usepackage{times}
\usepackage[margin=1in,top=0.45in]{geometry}
\usepackage{microtype}
\usepackage{amsmath,amssymb,amsfonts}
\usepackage{graphicx}
\graphicspath{{figures/}{logos/}}
\usepackage{booktabs,multirow,tabularx}
\usepackage[table]{xcolor}
\usepackage{caption}
\usepackage{enumitem}
\usepackage[most]{tcolorbox}
\usepackage{placeins}
\usepackage{wrapfig}
\usepackage{needspace}
\usepackage{titlesec}
\setlist{leftmargin=5mm}

\definecolor{ink}{HTML}{18334B}
\definecolor{tablehead}{HTML}{EAF0F7}
\definecolor{oursbg}{HTML}{E6F2EF}
\definecolor{gainbg}{HTML}{F3F8F6}
\newcolumntype{L}{>{\raggedright\arraybackslash}X}
\newcolumntype{R}{>{\raggedleft\arraybackslash}X}
\renewcommand{\arraystretch}{1.12}
\newcommand{\tableformat}{\small\setlength{\tabcolsep}{3.5pt}\renewcommand{\arraystretch}{1.12}}
\usepackage[authoryear,round]{natbib}
\setcitestyle{aysep={,},yysep={;}}
\usepackage{xurl}
\usepackage[colorlinks=true,linkcolor=cyan,citecolor=cyan,filecolor=magenta,urlcolor=blue]{hyperref}
\hypersetup{
  bookmarksdepth=2,
  pdftitle={UniMore: Fine-Grained Supervision for Unified Understanding and Generation},
  pdfauthor={Haoyue Bai, Yinyi Luo, Yicheng He, Qingsong Wen, Sharon Li, Jindong Wang}
}

\newcommand{\method}{UniMore}
\newcommand{\raw}{\theta_0}
\newcommand{\tgen}{\theta_{\mathrm{G}}}
\newcommand{\tcap}{\theta_{\mathrm{U}}}

\definecolor{abstractbg}{RGB}{230,242,250}
\definecolor{abstractborder}{RGB}{200,210,220}
\titleformat{\section}
  {\sffamily\bfseries\Large}
  {\thesection}
  {1em}
  {}
\renewenvironment{abstract}
{\begin{center}
\begin{tcolorbox}[
  colback=abstractbg,colframe=abstractborder,boxrule=0.6pt,arc=6pt,
  width=\textwidth,left=8pt,right=8pt,top=8pt,bottom=8pt
]}
{\end{tcolorbox}\end{center}}

\newcommand{\papertitle}{%
\sffamily\bfseries\fontsize{16}{19}\selectfont
UniMore: Fine-Grained Supervision for\\
Unified Understanding and Generation
}
\newcommand{\paperauthors}{%
\sffamily
Haoyue Bai$^1$, Yinyi Luo$^2$, Yicheng He$^3$,\\[3pt]
Qingsong Wen$^4$, Sharon Li$^5$, Jindong Wang$^{1,*}$
}
\newcommand{\paperaffiliations}{%
$^1$William \& Mary \quad $^2$Carnegie Mellon University\\[3pt]
$^3$University of Illinois Urbana-Champaign \quad $^4$Squirrel Ai Learning\\[3pt]
$^5$University of Wisconsin--Madison
}

\begin{document}
\thispagestyle{empty}

\noindent
\includegraphics[height=.52cm]{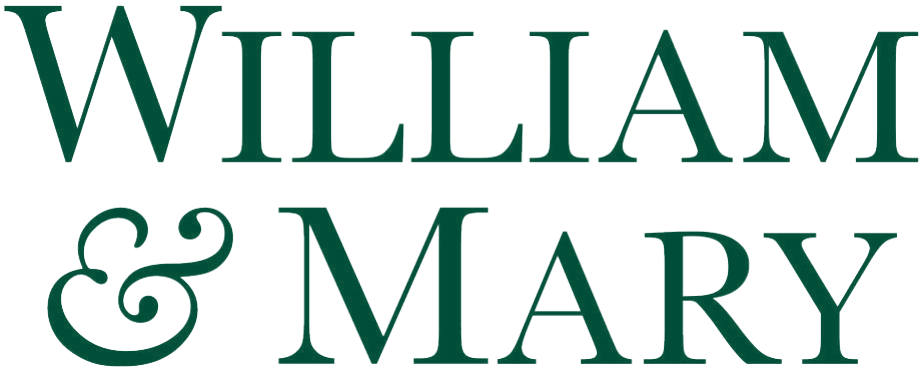}\hspace{0.38cm}%
\includegraphics[height=.52cm]{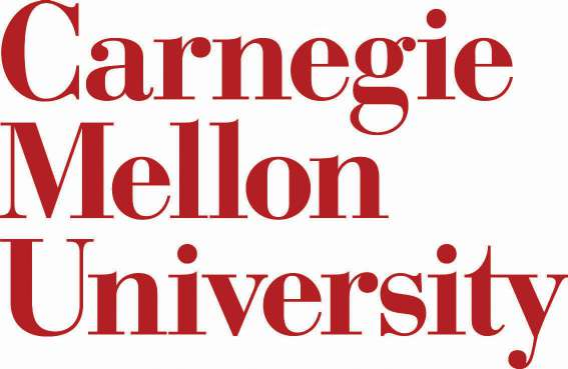}\hspace{0.38cm}%
\includegraphics[height=.42cm]{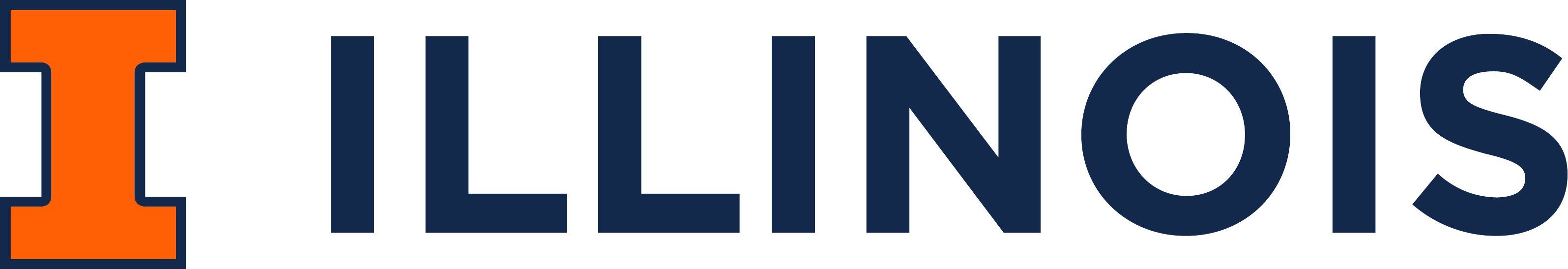}\hspace{0.38cm}%
\includegraphics[height=.52cm]{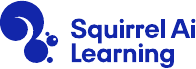}\hspace{0.38cm}%
\includegraphics[height=.52cm]{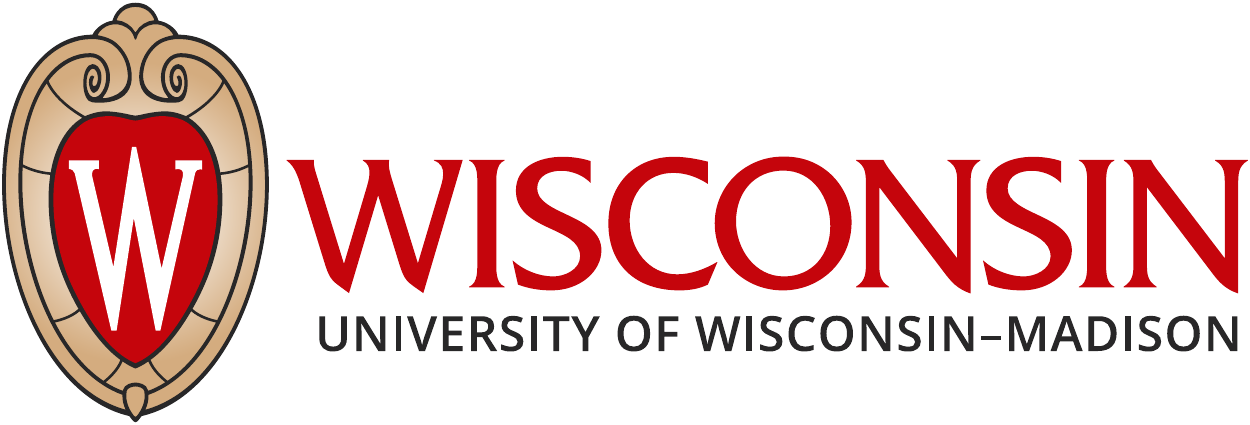}

\vspace{-.15in}
\noindent\rule{\textwidth}{0.8pt}
\vspace{0.32cm}
\begin{center}
  {\papertitle\par}
  \vspace{0.25cm}
  {\large\paperauthors\par}
  \vspace{0.2cm}
  {\normalsize\paperaffiliations\par}
  \vspace{0.15cm}
  {\small $^*$Corresponding author: \href{mailto:jdw@wm.edu}{jdw@wm.edu}\par}
\end{center}

\input{section/abstract}
\input{section/introduction}
\input{section/related_work}
\input{section/method}
\input{section/experiments}
\FloatBarrier
\input{section/conclusion}
\input{section/declarations}
\FloatBarrier
\setlength{\bibsep}{4pt plus 1pt minus 1pt}
\bibliographystyle{plainnat}
\bibliography{refs}

\end{document}

%% file: section/abstract.tex
\begin{abstract}
Unified multimodal models combine image understanding and generation within one system, creating an opportunity to improve both capabilities through paired image--text supervision. We examine the outputs of a pretrained unified model in both directions and observe a loss of specificity: generated images can contain the requested objects with incorrect attributes, while captions identify a scene without expressing its distinguishing details. These failures call for supervision that specifies both the visual content to realize and the descriptive differences to distinguish. We propose \method, a post-training method that derives these complementary learning tasks from the fine-grained content of image--text pairs. The generation route predicts grounded crops from their corresponding phrases, matching the scope of the condition and visual target while using the native visual objective. The understanding route combines positive-caption fitting with ranking against captions that omit or alter details, conditioned on the full image. The two routes are optimized independently, and their learned updates are combined into a single unified model without changing its inference procedure. Experiments show gains across multiple understanding and generation benchmarks. The code is available at \url{https://github.com/AIFrontierLab/TorchUMM/tree/main/src/umm/post_training/UniMore}.
\end{abstract}

%% file: section/introduction.tex
\section{Introduction}\label{sec:intro}

Unified multimodal models combine image understanding and generation within a common system~\citep{chameleon,emu3,transfusion}. Their shared architecture allows image--text pairs to support both visual understanding and image generation~\citep{bagel,januspro,showo2}. A shared architecture alone does not guarantee gains in both understanding and generation~\citep{luo2026beyond}. A central post-training question is how to organize the information in these paired observations into effective supervision for the two prediction directions. To examine where more focused supervision is needed, we inspect what a pretrained unified model preserves when it generates an image from a description and when it describes an image.

\begin{figure}[t]
\centering
\includegraphics[width=\textwidth]{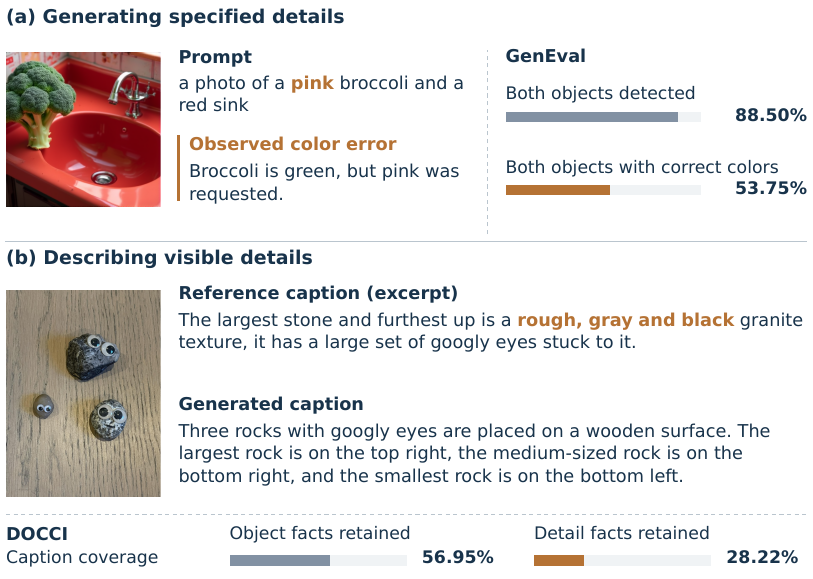}
\caption{\textbf{Object-level success can miss specific details.} Examples and aggregate diagnostics from pretrained BAGEL~\citep{bagel} compare \textbf{(a)} object detection with color fidelity on GenEval~\citep{geneval} and \textbf{(b)} object-fact with detail-fact retention in DOCCI captions~\citep{docci}.}\label{fig:motivation}
\end{figure}

Our diagnostics of pretrained BAGEL~\citep{bagel} reveal a gap between object-level success and the preservation of specific details (Figure~\ref{fig:motivation}). Given a request for \emph{pink broccoli} and a \emph{red sink}, the model generates both objects but leaves the broccoli green. When describing an image of three rocks with googly eyes, it identifies the objects and their arrangement but omits the largest rock's gray-and-black color and rough texture. These examples accompany a broader difference in the corresponding diagnostics. On GenEval's two-object color subset~\citep{geneval}, both objects are detected in 88.50\% of images, but only 53.75\% meet both color requirements. On DOCCI~\citep{docci}, free captions retain 56.95\% of reference object facts and 28.22\% of detail facts.

These failures motivate post-training that improves the realization and discrimination of specific visual content. Existing approaches strengthen the two capabilities through cross-task feedback~\citep{corl,unigame,srum}, reconstruction and latent alignment~\citep{reca,latentumm}, or adaptive coordination during reasoning~\citep{unipath}. We focus on how the correspondences already present in paired observations can specify the visual content to predict and the descriptive differences to distinguish. For generation, a phrase and its grounded region define a prediction task with a matching scope. For understanding, captions that share scene content but omit or alter a detail specify which differences should affect the model's preference.

We introduce \textit{\method} to turn these two uses of paired observations into post-training tasks. The generation route pairs a caption phrase with its grounded crop and trains the backbone with its native visual prediction objective. The understanding route combines positive-caption likelihood with ranking against detail-deleted, detail-corrupted, and unrelated captions under full-image conditioning. Both routes start from the same pretrained model and are optimized independently. Static task-vector addition~\citep{taskarith} combines their updates into a single unified model. The method reorganizes supervision from paired data while using the model's existing prediction interfaces for visual and textual outputs.

Extensive experiments on BAGEL~\citep{bagel}, Janus-Pro~\citep{januspro}, and Show-o2~\citep{showo2} demonstrate the effectiveness of \method\ across understanding and generation benchmarks and targeted detail diagnostics.

Our contributions are:
\begin{itemize}
\item \textbf{A diagnosis of detail failures.} We connect concrete generation and captioning errors to the gap between object-level success and the preservation of specific semantic content.
\item \textbf{Supervision construction from paired data.} We construct grounded phrase-to-crop prediction and full-image caption discrimination from paired data, then compose the adaptations in one model.
\item \textbf{Evaluation across prediction interfaces.} Experiments across three backbones show improvements on understanding and generation benchmarks, with target comparisons and paired diagnostics examining the effects of the supervision design.
\end{itemize}

%% file: section/related_work.tex
\section{Related Work}\label{sec:related}

\subsection{Unified multimodal architectures}
Unified systems differ in how they connect language and visual prediction. NExT-GPT~\citep{nextgpt} connects a language model to modality-specific encoders and decoders through learned adapters. VITRON~\citep{vitron} combines regional visual inputs with specialist modules for generation, segmentation, and editing. X-Fusion~\citep{xfusion} instead introduces modality-specific weights around a frozen language model to support understanding and generation.

Native unified models also differ in their visual representations and prediction objectives. Chameleon and Emu3 use autoregressive prediction over discrete multimodal tokens~\citep{chameleon,emu3}. Tar~\citep{tar} aligns visual tokens with the language vocabulary through a text-aligned codebook. Show-o combines autoregressive text modeling with discrete image diffusion~\citep{showo}, while Transfusion couples next-token text prediction with diffusion over continuous image representations~\citep{transfusion}. FlexUMM~\citep{luo2026flexumm} dynamically allocates shared and specialized computation and adjusts processing depth across inputs. Among the backbones studied here, BAGEL~\citep{bagel} and Show-o2~\citep{showo2} support continuous image generation, while Janus-Pro~\citep{januspro} separates visual encoding for understanding and generation within an autoregressive model. This diversity motivates post-training targets that can be expressed through each backbone's native prediction interface for visual and textual outputs.

\subsection{Post-training and capability coordination}
One post-training approach uses cross-task feedback to guide optimization. CoRL~\citep{corl} and UniGame~\citep{unigame} formulate this interaction through co-reinforcement learning and adversarial training, respectively. UniRL~\citep{unirl} uses generated images to train understanding and understanding outputs to supervise generation in a self-improvement loop. GvU~\citep{gvu} and SRUM~\citep{srum} use understanding-derived rewards to guide generation. Another approach connects the tasks through their representations. RecA~\citep{reca} reconstructs an image from its visual-understanding embeddings. UNO~\citep{uno} passes noised generative representations to a frozen understanding expert, using captioning and visual-regression objectives to supervise the generator.

LatentUMM~\citep{latentumm} explicitly aligns cross-modal representations and generation--re-encoding transformations, then stabilizes latent trajectories through stochastic rollouts and preference optimization. Coordination can also occur during reasoning. UniPath~\citep{unipath} uses a lightweight planner to select an input-dependent coordination path and an executor trained on role-aligned trajectories to carry it out. AD-Loop~\citep{adloop} learns to interleave textual analysis and visual drafting during problem solving. Controlled studies of native unified models further show that transfer depends on computational specialization and shared task knowledge~\citep{native_synergy}. \method\ studies the supervision targets used during adaptation, combining grounded visual prediction with full-image caption discrimination.

\subsection{Fine-grained vision-language supervision}
Fine-grained image--text matching requires sensitivity to how objects, attributes, and relations are composed. Winoground probes this ability using image pairs and captions with the same words in different orders~\citep{winoground}. ARO evaluates attribute, relation, and order sensitivity, and NegCLIP augments contrastive fine-tuning with composition-aware hard negatives~\citep{negclip}. Negative-caption quality also affects what such comparisons measure. SugarCrepe shows that linguistic artifacts can allow correct choices without grounding the decision in the relevant visual evidence~\citep{sugarcrepe}.

Regional correspondence and detail-sensitive text supervision provide a foundation for learning beyond whole-image semantics. GLaMM~\citep{glamm} links natural-language responses to pixel-level segmentation masks, while GenerateU~\citep{generateu} learns to translate detected visual regions into free-form object names. CLIP-based caption rewards encourage descriptions to express distinctive visual content~\citep{clipreward}. FG-CLIP~\citep{fgclip} combines global and regional feature alignment with hard negative descriptions that alter local attributes. TokenCompose~\citep{tokencompose} uses segmentation masks to supervise token-level cross-attention in text-to-image diffusion. These methods make fine-grained correspondence explicit through region--language grounding, representation alignment, and supervision of the text-to-image generation process.

\method\ places detail constraints in the native prediction targets of a unified model. For generation, a caption phrase conditions visual prediction with its grounded image crop as the output target. For understanding, the full image conditions both a positive caption and alternatives that change specific details, and their conditional likelihoods define the ranking objective. This formulation gives the two directions distinct responsibilities: predicting the visual evidence associated with a phrase and distinguishing descriptions that differ in how they express that evidence.

%% file: section/method.tex
\section{Method}\label{sec:method}

\subsection{Problem formulation and overview}
We are given a pretrained unified multimodal model with parameters $\raw$ and a collection of paired images and captions $\mathcal D=\{(x_i,y_i)\}_{i=1}^{N}$, where $x_i$ is an image, $y_i$ is its description, and $N$ is the number of pairs. The model supports image generation conditioned on text and text prediction conditioned on an image. Our goal is to adapt it so that generated images better realize the details specified in language and textual predictions better distinguish the details supported by visual evidence, while retaining both capabilities in one model.

\begin{figure}[t]
\centering
\includegraphics[width=\textwidth]{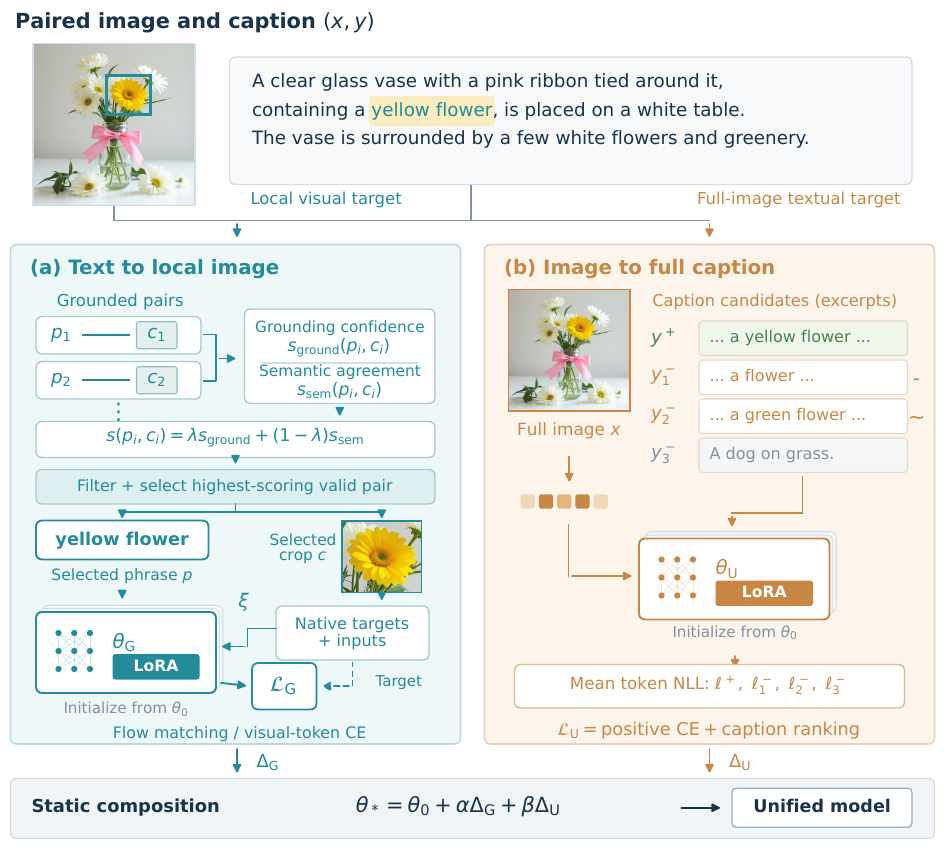}
\caption{\textbf{Overview of \method.} An image--caption pair supports \textbf{(a)} scored phrase--crop selection for visual prediction and \textbf{(b)} full-image caption fitting and ranking against detail-deleted, detail-corrupted, and unrelated alternatives. Both routes adapt independently from the same backbone, then combine their updates into one model. Candidate cards are schematic, and captions are excerpted.}\label{fig:method}
\end{figure}

\method\ turns specific content in paired images and captions into visual prediction and textual comparison tasks through the backbone's native prediction interfaces (Figure~\ref{fig:method}). For generation, the model learns to predict an image crop $c$ from its corresponding phrase $p$, matching the scope of the text condition and visual target. For understanding, it uses the full image $x$ to score a positive caption $y^+$ and alternatives that omit or change its details, making their relative preference part of the supervision. We use the resulting generation and understanding datasets, $\mathcal D_{\mathrm G}$ and $\mathcal D_{\mathrm U}$, to adapt two copies of the same pretrained model independently. Their parameter updates are then composed into the final model $\theta_*$.

\subsection{Grounded targets for local visual realization}\label{sec:generation}
The generation route uses a phrase--region correspondence within an image--caption pair to define a local prediction task. Whole-image prediction supervises an entire scene, including background and objects beyond a particular phrase. We pair that phrase with its grounded crop so that the text condition and visual target refer to the same local content.
To construct this supervision, we first locate candidate regions for descriptive phrases and select pairs that agree in both location and semantic content. We then train the model to predict each retained crop from its phrase using the backbone's native visual objective.

\paragraph{Constructing reliable local targets.}
We construct supervised phrase--crop pairs, using each text phrase as the input condition and a corresponding region cropped from the source image as the visual target. The selected crop must contain the named object and match the description used to predict it. We therefore begin by extracting object phrases together with their modifiers and grounding them in the paired image. We then assess each candidate through its detection confidence and its semantic agreement with the phrase.

For grounding confidence, we use a frozen text-conditioned detector, GroundingDINO~\citep{groundingdino}, with the full image and the extracted phrases as its inputs. Let $a_{c,k}$ be its matching logit between the box used to obtain crop $c$ and text-token position $k$. After matching the box's decoded text label to phrase $p$, we retain its detector confidence. With $\sigma$ denoting the sigmoid function, this score is
\begin{equation}
 s_{\mathrm{ground}}(p,c)=\max_k\sigma(a_{c,k}),
 \label{eq:ground_confidence}
\end{equation}
where the maximum spans all detector text-token positions in the joint phrase query. The score is retained from the proposed box when constructing the corresponding crop from the full source image.

To evaluate the cropped content against the complete phrase, we encode them with the frozen image and text encoders of CLIP~\citep{clip}. Let $\mathbf v_c$ and $\mathbf t_p$ denote their visual and textual feature vectors. Semantic agreement is their cosine similarity rescaled to $[0,1]$:
\begin{equation}
 s_{\mathrm{sem}}(p,c)=\frac{1}{2}\left(
 1+\frac{\mathbf v_c^{\top}\mathbf t_p}
 {\|\mathbf v_c\|_2\|\mathbf t_p\|_2}\right).
 \label{eq:semantic_agreement}
\end{equation}
This crop--phrase comparison complements the detector's response to the original image. Both scores lie in $[0,1]$, and we combine them as
\begin{equation}
 s(p,c)=\lambda s_{\mathrm{ground}}(p,c)+(1-\lambda)s_{\mathrm{sem}}(p,c).
 \label{eq:ground_score}
\end{equation}
The weight $\lambda\in[0,1]$ balances localization confidence and semantic agreement. After confidence and geometric filtering, we retain the highest-scoring valid phrase--crop match from each source pair. Pairs without a suitable visual target are excluded. The resulting dataset $\mathcal D_{\mathrm G}=\{(p,c)\}$ uses the phrase as the text condition and the selected crop as the complete visual target.

\paragraph{Learning to realize the phrase.}
With these pairs in place, we train the model to reproduce the selected visual content from its description. We retain the backbone's native visual objective, so the same target construction can be used with different visual representations. For a flow-matching generator, let $z_0=E(c)$ be the crop encoded by the visual encoder $E$, let $\epsilon\sim\mathcal N(0,I)$ be Gaussian noise, and let $z_t=(1-t)z_0+t\epsilon$ be the noised latent at interpolation time $t$. Under this data-to-noise convention, the target velocity is $\epsilon-z_0$, and the phrase-conditioned velocity predictor $v_{\theta}$ is trained with
\begin{equation}
 \mathcal L_{\mathrm G}^{\mathrm{flow}}=
 \mathbb E_{(p,c),t,\epsilon}\left[
 \operatorname{MSE}\!\left(v_{\theta}(z_t,t,p),\epsilon-z_0\right)
 \right].\label{eq:flow}
\end{equation}
The mean squared error averages over supervised latent positions and channels. Other continuous generators retain their native transport convention and time sampling. For an autoregressive visual generator, the crop is encoded as $M$ visual tokens $q_1,\ldots,q_M$, and the objective is
\begin{equation}
 \mathcal L_{\mathrm G}^{\mathrm{AR}}
 =-\frac{1}{M}\sum_{j=1}^{M}\log p_{\theta}(q_j\mid p,q_{<j}).\label{eq:ar}
\end{equation}
Here $q_{<j}$ is the preceding visual-token sequence. The native visual input $\xi$ in Figure~\ref{fig:method} denotes the noised latent and time $(z_t,t)$ for continuous prediction, or the token prefix $q_{<j}$ for autoregressive prediction. In both forms, the phrase specifies what to generate and the grounded crop supplies the matching visual content that the model learns to predict.

\subsection{Full-context targets for caption discrimination}\label{sec:understanding}
The understanding route uses the specific content of a paired caption to define comparisons between related descriptions. Positive-caption fitting teaches the model to reproduce a reference description, but it does not explicitly require that description to outrank alternatives with missing or altered details. We retain this direct supervision and add a ranking constraint that makes the relative preference explicit. To do so, we first construct caption alternatives, then score all candidates using the full image, and jointly optimize positive-caption likelihood and their separation. The full image preserves the scene context needed to judge object attributes and relationships between objects.

\paragraph{Constructing informative comparisons.}
To specify which content the model should distinguish, we pair the positive caption $y^+$ with alternatives from three sources. \emph{Detail deletion} removes specific content, \emph{detail corruption} substitutes a detail, and \emph{unrelated text} provides a coarse semantic contrast. The first two retain shared content with the reference, allowing the comparison to focus on what has been omitted or changed. A deleted caption can remain true, but the objective favors the fuller reference description. Thus, the target distinction concerns both fidelity to the image and completeness relative to the supplied reference. The resulting image--caption groups form the understanding dataset $\mathcal D_{\mathrm U}$ used for caption discrimination under full-image conditioning.

\paragraph{Learning caption preferences.}
Ranking supplements positive-caption fitting with an explicit preference over related descriptions. We use a margin-based hinge loss so that a comparison stops contributing to the ranking penalty once the required separation is reached. This is useful for incomplete descriptions that may still be valid, since they need to rank below the reference without being continually suppressed. We score the candidates using the model's own caption likelihood, keeping the supervision within its existing text-prediction interface. To compare candidates of different lengths, we use the mean negative log likelihood over the $|y|$ supervised caption tokens, conditioned on the full image:
\begin{equation}
 \ell_{\theta}(y\mid x)=-\frac{1}{|y|}\sum_{j=1}^{|y|}
 \log p_{\theta}(y_j\mid x,y_{<j}).\label{eq:caption}
\end{equation}
Here $y_{<j}$ denotes the preceding caption tokens, and prompt and image tokens are excluded from the loss. Length normalization removes the direct sum-over-tokens effect. For training example $i$, let $\mathcal N_i$ index its available negative captions $y_k^-$. We combine positive-caption cross entropy with a hinge ranking term:
\begin{equation}
\begin{aligned}
 \mathcal L_{\mathrm U}
 ={}&\ell_{\theta}(y^+\mid x)+\frac{1}{|\mathcal N_i|}\sum_{k\in\mathcal N_i}\!\!
 \\[-2pt]
 &\left[m+\ell_{\theta}(y^+\mid x)-\ell_{\theta}(y_k^-\mid x)\right]_+.
\end{aligned}\label{eq:ranking}
\end{equation}
The margin $m>0$ specifies the desired separation, and $[a]_+=\max(0,a)$. The first term fits the reference caption, while the second penalizes alternatives whose mean loss is less than $m$ above the mean token loss of the reference caption.

\subsection{Independent adaptation and static composition}\label{sec:merge}
The generation and understanding objectives update shared backbone parameters but need not favor the same changes. In alternating training, an update that improves local visual prediction can increase caption-ranking loss, or vice versa, partially undoing progress on the other objective. Such conflicting updates are a known source of interference in multi-task optimization~\citep{pcgrad}. We therefore train the two routes independently from the same initialization, keeping one route's updates from overwriting the other's during training. After adaptation, we use task-vector addition~\citep{taskarith} to combine the learned updates into a single unified model. Separate composition weights control their contributions and can be adjusted without retraining either route.
Let $\tgen$ and $\tcap$ be the parameters learned from $\mathcal D_{\mathrm G}$ and $\mathcal D_{\mathrm U}$, respectively. Their updates relative to the shared initialization are
\begin{equation}
 \Delta_{\mathrm G}=\tgen-\raw,\qquad
 \Delta_{\mathrm U}=\tcap-\raw.
\end{equation}
The final model is obtained by linear composition:
\begin{equation}
 \theta_* = \raw+\alpha\Delta_{\mathrm G}+\beta\Delta_{\mathrm U}.
 \label{eq:merge}
\end{equation}
The scalar weights $\alpha$ and $\beta$ control the contributions of the two updates and accommodate differences in their scales. They are calibrated empirically and fixed for the evaluated models. No post-merge training is performed. Composition preserves the backbone's parameter shapes, so the final model serves both tasks through its original inference interfaces.

%% file: section/experiments.tex
\section{Experiments}\label{sec:experiments}
\subsection{Experimental setup}\label{sec:setup}

\paragraph{Implementation and baselines.}
Our implementation builds on TorchUMM~\citep{torchumm}, using its model interfaces and benchmark suite. The main comparisons include native unified multimodal models and post-training alternatives based on BAGEL~\citep{bagel}. SFT provides a supervised fine-tuning reference. RecA~\citep{reca} trains image reconstruction from understanding embeddings, while UniGame~\citep{unigame} uses adversarial perturbations at the shared token interface to challenge understanding. IRG~\citep{irg} alternates textual reasoning and reflection with image generation. LatentUMM~\citep{latentumm} aligns cross-modal representations and generation--re-encoding transformations through dual latent alignment.

\paragraph{Backbones and training data.}
We evaluate BAGEL-7B-MoT, Janus-Pro-7B~\citep{januspro}, and Show-o2-7B~\citep{showo2}. Training pairs are drawn from \texttt{text-to-image-2M}~\citep{zou2026advancing}. The generation dataset contains 2,460 phrase--crop pairs retained from 2,500 source pairs, and the caption-discrimination dataset contains 4,000 image--caption pairs.

\paragraph{Phrase--crop construction.}
Lexical and regular-expression rules extract predefined object nouns and nearby modifiers, producing phrases of one to seven words and at most eight candidates per image. GroundingDINO~\citep{groundingdino} proposes regions with box and text thresholds of 0.25 and 0.20. CLIP ViT-B/32~\citep{clip} supplies semantic agreement, with cosine similarity mapped to $(1+\mathrm{cos})/2$. Equation~\ref{eq:ground_score} uses $\lambda=0.65$, assigning weights of 0.65 and 0.35 to grounding confidence and semantic agreement. The final score threshold is 0.25. Boxes are padded by 0.08 and filtered to occupy between 0.01 and 0.80 of the image area, after which the highest-scoring accepted match is retained. Pairs dominated by text or logos, and pairs without a valid grounded region, are excluded.

\paragraph{Caption alternatives.}
Lexical rules construct and deduplicate up to three alternatives per positive caption. Detail deletion removes attribute, count, relation, or object-noun terms. Fallback candidates retain a first sentence or approximately the first 60\% of a sufficiently long caption. Detail corruption uses fixed substitutions such as red/blue, two/three, and left/right. Sampling is seeded by the example identity, and selection cycles through deletion, corruption, and unrelated text with deduplication and random-caption fallback. The ranking margin in Equation~\ref{eq:ranking} is $m=0.2$.

\paragraph{Backbone-specific implementation.}
BAGEL uses $256\times256$ crops with the flow objective in Equation~\ref{eq:flow}. Janus-Pro uses the visual-token objective in Equation~\ref{eq:ar}, while Show-o2 retains its native continuous visual loss, transport direction, and time sampling. Both retain their native preprocessing. Show-o2 caption targets are limited to 220 tokens, including the end-of-sequence token.

\Needspace{16\baselineskip}
\begin{wraptable}{r}{0.54\textwidth}
\centering
\captionsetup{justification=raggedright,singlelinecheck=false}
\caption{\textbf{Training settings.} G/U denote generation/understanding.}\label{tab:training}
\tableformat
\footnotesize
\setlength{\tabcolsep}{2.2pt}
\begin{tabularx}{\linewidth}{Lcrrrr}
\toprule
\rowcolor{tablehead}
\textbf{Backbone} & \textbf{G/U} & \textbf{LR} & \textbf{Steps} & \textbf{Warm.} & \textbf{Procs}\\
\midrule
\multirow{2}{*}{BAGEL} & G & $5\!\times\!10^{-5}$ & 1,000 & 50 & 4\\
 & U & $1.8\!\times\!10^{-5}$ & 600 & 60 & 2\\
\addlinespace[3pt]
\multirow{2}{*}{Janus-Pro} & G & $1\!\times\!10^{-5}$ & 250 & 25 & 1\\
 & U & $1.8\!\times\!10^{-5}$ & 600 & 60 & 1\\
\addlinespace[3pt]
\multirow{2}{*}{Show-o2} & G & $5\!\times\!10^{-6}$ & 500 & 50 & 1\\
 & U & $1.8\!\times\!10^{-5}$ & 300 & 30 & 1\\
\bottomrule
\end{tabularx}
\end{wraptable}

\paragraph{Optimization.}
All routes use LoRA~\citep{lora} with rank 16, alpha 32, and dropout 0.05. Learning rates (LR), warmup steps, and training processes (Procs) are listed in Table~\ref{tab:training}, while visual encoders or tokenizers remain fixed. The implementations use AdamW with momentum coefficients $(0.9,0.95)$, an epsilon of $10^{-15}$, zero weight decay, gradient norm clipping at 1.0, and a constant learning rate after linear warmup. For Janus-Pro and Show-o2, generation and caption adaptation accumulate four and two examples per optimizer step, respectively. BAGEL uses packed token batches, with expected/maximum token budgets of 256/512 for generation and 768/1,536 for understanding. In addition to LoRA, BAGEL trains the generation bridge and time components or the visual connector for the corresponding route. Janus-Pro and Show-o2 adapt language-model attention and MLP projections through LoRA.

For BAGEL, Equation~\ref{eq:ranking} is evaluated only for groups containing a positive caption and at least one associated negative in the same packed batch. Thus, $\mathcal N_i$ contains the negatives available with that positive in the batch. Groups lacking either contribute neither the positive-caption term nor the ranking term. Janus-Pro and Show-o2 keep each positive caption and its associated negatives together when computing the ranking objective for each image--caption group.

\paragraph{Evaluation protocol.}
Adapted variants share evaluation data, inference settings, and scorers within each controlled comparison. Paired diagnostics also evaluate the corresponding pretrained model under the same settings. Our BAGEL generation uses $512\times512$ images, 50 sampling steps, and seed 42. Janus-Pro and Show-o2 retain their native generation sizes of 384 and 432 pixels, respectively, so comparisons are made within each architecture. WISE~\citep{wise} uses Qwen2.5-VL-72B-Instruct~\citep{qwen25vl} to evaluate the generated images.

Understanding benchmarks comprise MMMU~\citep{mmmu} (900 examples), MathVista~\citep{mathvista} (1,000), MMStar~\citep{mmstar} (1,500), MMBench-EN, and MM-Vet~\citep{mmvet}. For Janus-Pro and Show-o2, MMBench uses TEST EN V11 with 7,299 examples and the official aggregation.

Generation benchmarks comprise GenEval (553 prompts, four images each), WISE (1,000 prompts), and UEval~\citep{ueval} (1,000 questions). GenEval and WISE evaluate text-to-image generation, while UEval evaluates instruction following in joint image--text generation. RealUnify~\citep{realunify} and Unison~\citep{unison} evaluate cross-task interaction and consistency between understanding and generation, with task definitions and evaluation sizes given in Section~\ref{sec:consistency}. The detail-diagnostic protocols are specified alongside their results in Sections~\ref{sec:detail_diagnostics} and~\ref{sec:detail_results}.

Benchmark scores use a 0--100 scale, and ``--'' in tables denotes an unavailable result.

\paragraph{Model composition.}
All three backbones use $\alpha=1.0$ and $\beta=0.02$ in Equation~\ref{eq:merge}. We select $\beta$ on BAGEL using accuracy on 500 MMBench-Dev~\citep{mmbench} questions and the GenEval~\citep{geneval} macro-average score on 100 prompts (Section~\ref{sec:ablations}). BAGEL adds parameter updates sequentially in BF16. For Janus-Pro and Show-o2, updates are accumulated in FP32 before casting to the model precision. No additional training follows composition of the two route-specific updates.

\subsection{Main understanding and generation results}
In Table~\ref{tab:understanding_main}, MathVista includes multiple-choice (MC), free-form (FF), and overall (All) scores. Avg. is the unweighted mean of the five benchmarks, using MathVista All. Table~\ref{tab:generation_main} reports overall UEval scores and the six WISE categories together with their overall score (All).

\begin{table}[!htbp]
\centering
\caption{\textbf{Multimodal understanding.} Bold and underline mark the best and second-best results within the BAGEL post-training group.}\label{tab:understanding_main}
\tableformat
\fontsize{9.5}{11}\selectfont
\renewcommand{\arraystretch}{1.06}
\setlength{\tabcolsep}{3.5pt}
\begin{tabularx}{\linewidth}{Lcrrrrrrrr}
\toprule
\rowcolor{tablehead}
 & & & & & & \multicolumn{3}{c}{\textbf{MathVista}} & \\
\rowcolor{tablehead}
\multirow{-2}{*}{\textbf{Method}} & \multirow{-2}{*}{\textbf{Size}} & \multirow{-2}{*}{\textbf{MMMU}} & \multirow{-2}{*}{\textbf{MMBench}} & \multirow{-2}{*}{\textbf{MM-Vet}} & \multirow{-2}{*}{\textbf{MMStar}} & MC & FF & All & \multirow{-2}{*}{\textbf{Avg.}}\\
\midrule
\multicolumn{10}{l}{\textit{Native unified multimodal models}}\\
BAGEL~\citeyearpar{bagel} & 7B & 51.90 & 82.65 & 65.90 & 63.20 & 80.19 & 61.52 & 71.60 & 67.05\\
Janus-Pro~\citeyearpar{januspro} & 7B & 40.70 & 67.64 & 33.70 & 42.00 & 51.30 & 32.83 & 42.80 & 45.37\\
Janus~\citeyearpar{janus} & 1.3B & 27.30 & 27.69 & 27.00 & 26.67 & 35.37 & 16.30 & 26.60 & 27.05\\
JanusFlow~\citeyearpar{janusflow} & 1.3B & 29.00 & 64.16 & 31.80 & 39.19 & 39.07 & 29.78 & 34.80 & 39.79\\
Show-o2~\citeyearpar{showo2} & 7B & 47.90 & 77.25 & 47.10 & 54.53 & 63.52 & 37.39 & 51.50 & 55.66\\
Show-o2~\citeyearpar{showo2} & 1.5B & 37.10 & 65.09 & 46.10 & 43.87 & 53.33 & 19.78 & 37.90 & 46.01\\
Show-o~\citeyearpar{showo} & 1.3B & 26.10 & 43.48 & 23.30 & 27.87 & 44.07 & 11.30 & 29.00 & 29.95\\
Emu3~\citeyearpar{emu3} & 8B & 31.40 & 59.53 & 30.00 & 43.60 & 57.59 & 30.00 & 44.90 & 41.89\\
OmniGen2~\citeyearpar{omnigen2} & 7B & 46.00 & 76.36 & 62.70 & 52.47 & 71.67 & 53.91 & 63.50 & 60.21\\
MMaDA~\citeyearpar{mmada} & 8B & 28.90 & 28.16 & 11.40 & 31.73 & 38.70 & 8.70 & 24.90 & 25.02\\
Ovis-U1~\citeyearpar{ovisu1} & 3.6B & 43.70 & 78.55 & 62.80 & 58.27 & 75.74 & 60.00 & 68.50 & 62.36\\
Emu3.5~\citeyearpar{emu35} & 34B & 29.20 & 17.56 & 28.00 & 30.27 & 41.67 & 17.61 & 30.60 & 27.13\\
\midrule
\multicolumn{10}{l}{\textit{BAGEL post-training}}\\
SFT & 7B & 52.60 & \underline{82.65} & 61.20 & 68.27 & 80.00 & \underline{65.00} & 73.10 & 67.56\\
RecA~\citeyearpar{reca} & 7B & 52.30 & \underline{82.65} & 66.10 & 68.47 & 79.81 & 64.57 & 72.80 & \underline{68.46}\\
UniGame~\citeyearpar{unigame} & 7B & 52.40 & \textbf{82.75} & 60.70 & \underline{68.73} & 79.63 & 63.48 & 72.20 & 67.36\\
IRG~\citeyearpar{irg} & 7B & 48.00 & 60.47 & 40.70 & 61.53 & 73.52 & 61.52 & 68.00 & 55.74\\
LatentUMM~\citeyearpar{latentumm} & 7B & \underline{53.20} & 81.04 & \underline{67.20} & 65.33 & \underline{80.37} & \textbf{65.65} & \underline{73.60} & 68.07\\
\addlinespace[3pt]
\rowcolor{oursbg}
\textbf{\method} & 7B & \textbf{53.30} & \textbf{82.75} & \textbf{67.30} & \textbf{70.13} & \textbf{80.74} & \textbf{65.65} & \textbf{73.80} & \textbf{69.46}\\
\bottomrule
\end{tabularx}
\end{table}

\paragraph{Understanding performance.}
We first test whether detail-targeted adaptation improves general visual understanding. Table~\ref{tab:understanding_main} compares native unified models and BAGEL post-training methods on five benchmarks covering perception and reasoning. \method\ achieves the highest scores among the BAGEL post-training methods on MMMU, MM-Vet, MMStar, and MathVista. The clearest improvement over the base model is on MMStar, which rises from 63.20 to 70.13. MathVista also improves by 2.20 points, with gains in both multiple-choice and free-form questions. These results show that the benefits extend to answering visual questions, consistent with the goal of learning more discriminative image--text correspondences through caption supervision.

\begin{table}[!htbp]
\centering
\caption{\textbf{Generation performance.} Bold and underline mark the best and second-best results within the BAGEL post-training group.}\label{tab:generation_main}
\tableformat
\fontsize{9.5}{11}\selectfont
\renewcommand{\arraystretch}{1.06}
\setlength{\tabcolsep}{8pt}
\begin{tabularx}{\linewidth}{Lrrrrrrrr}
\toprule
\rowcolor{tablehead}
 & & \multicolumn{7}{c}{\textbf{WISE}}\\
\rowcolor{tablehead}
\multirow{-2}{*}{\textbf{Method}} & \multirow{-2}{*}{\textbf{UEval}} & Culture & Time & Space & Bio. & Phys. & Chem. & All\\
\midrule
\multicolumn{9}{l}{\textit{Native unified multimodal models}}\\
BAGEL & 30.90 & 38.83 & 43.86 & 47.14 & 36.20 & 42.05 & 29.40 & 39.89\\
Janus-Pro & 20.60 & 36.16 & 38.53 & 47.89 & 36.05 & 47.45 & 24.85 & 38.11\\
Janus & -- & 20.80 & 27.07 & 35.08 & 17.05 & 19.10 & 10.95 & 22.22\\
JanusFlow & -- & 27.31 & 32.22 & 39.47 & 32.15 & 28.60 & 19.05 & 29.64\\
Show-o2 (7B) & 29.10 & 36.41 & 34.97 & 45.19 & 34.55 & 36.90 & 23.90 & 35.95\\
Show-o2 (1.5B) & 22.10 & 31.11 & 35.63 & 43.57 & 31.50 & 38.40 & 23.10 & 33.49\\
Show-o & -- & 28.65 & 32.25 & 41.32 & 27.50 & 33.00 & 19.80 & 30.37\\
Emu3 & -- & 34.63 & 34.82 & 37.11 & 33.10 & 36.85 & 21.30 & 33.73\\
OmniGen2 & 25.80 & 41.80 & 40.42 & 48.87 & 36.35 & 38.75 & 28.10 & 40.29\\
MMaDA & 4.40 & 65.02 & 68.14 & 74.92 & 66.20 & 74.20 & 42.05 & 65.60\\
Ovis-U1 & 34.20 & 36.43 & 39.91 & 48.31 & 34.05 & 40.90 & 23.90 & 37.55\\
Emu3.5 & -- & 70.01 & 56.83 & 69.44 & 64.35 & 60.85 & 40.60 & 63.31\\
BLIP3-o~\citeyearpar{blip3o} & -- & 40.28 & 41.86 & 52.59 & 40.25 & 42.55 & 30.00 & 41.38\\
TokenFlow~\citeyearpar{tokenflow} & -- & 32.53 & 36.26 & 33.57 & 29.15 & 26.05 & 15.10 & 30.56\\
DeepGen~\citeyearpar{deepgen} & -- & 59.89 & 49.55 & 61.02 & 47.65 & 55.15 & 40.80 & 54.70\\
\midrule
\multicolumn{9}{l}{\textit{BAGEL post-training}}\\
SFT & 31.40 & 22.01 & 21.17 & 34.66 & 17.30 & 25.45 & 15.10 & 22.74\\
RecA & 31.00 & 40.35 & 41.47 & \underline{54.32} & \underline{39.85} & 46.30 & \textbf{33.40} & \underline{42.25}\\
UniGame & 31.00 & 39.56 & 41.38 & 48.76 & 35.90 & 43.55 & 31.55 & 40.32\\
IRG & 9.10 & 36.74 & 40.81 & 46.50 & 35.75 & 44.95 & 26.55 & 38.42\\
LatentUMM & \textbf{31.80} & \underline{41.31} & \underline{41.55} & \textbf{55.47} & 35.90 & \underline{46.95} & 32.06 & 41.80\\
\addlinespace[3pt]
\rowcolor{oursbg}
\textbf{\method} & \underline{31.60} & \textbf{42.86} & \textbf{43.89} & 52.26 & \textbf{41.15} & \textbf{47.00} & \underline{32.40} & \textbf{43.48}\\
\bottomrule
\end{tabularx}
\end{table}

\paragraph{Generation performance.}
We next examine whether learning from local phrase--crop pairs benefits complete-image generation and joint image--text generation. Table~\ref{tab:generation_main} evaluates WISE and UEval under their standard protocols. On WISE, \method\ improves BAGEL from 39.89 to 43.48 and outperforms the strongest post-training baseline, RecA, by 1.23 points. It improves all six categories over BAGEL and leads the post-training group in four. The accompanying UEval improvement extends the gains to instruction following in responses containing both images and text. Together, these results suggest that supervision concentrated on local visual content can transfer to broader generation tasks.

\FloatBarrier
\Needspace{19\baselineskip}
\subsection{Detail retention and realization}\label{sec:detail_diagnostics}

\begin{wrapfigure}{r}{0.49\textwidth}
\centering
\captionsetup{justification=raggedright,singlelinecheck=false}
\includegraphics[width=\linewidth]{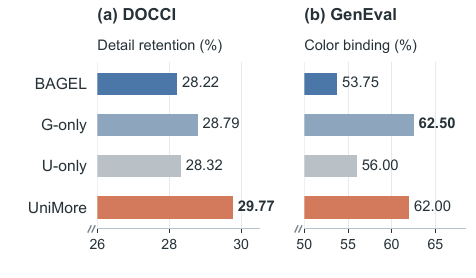}
\caption{\textbf{Detail retention and realization on BAGEL.} Horizontal axes are truncated. Bold labels mark the highest score in each panel.}\label{fig:detail_diagnostics}
\end{wrapfigure}

We revisit the diagnostics in Figure~\ref{fig:motivation} to examine whether the benchmark gains also address the observed detail failures. Figure~\ref{fig:detail_diagnostics} compares pretrained BAGEL, generation-only adaptation (G-only), understanding-only adaptation (U-only), and \method\ under common evaluation settings. DOCCI~\citep{docci} measures the detail facts retained in free captions, while GenEval tests whether both requested objects have their specified colors in the same generated image.

The caption-coverage diagnostic in Figures~\ref{fig:motivation} and~\ref{fig:detail_diagnostics} uses 500 images from the official DOCCI test set, a common free-description prompt, greedy decoding, and a 500-token limit. A Qwen3-32B judge~\citep{qwen3} scores reference facts derived from the dataset's Davidsonian Scene Graph questions~\citep{dsg} and checked against the human descriptions. The inventory contains 22,620 supported facts, including 13,578 detail facts. Coverage is the fraction of reference facts expressed in a generated caption and does not measure recognition when the model is explicitly asked about individual visual attributes.

\method\ improves detail retention by 1.55 percentage points and color-binding accuracy by 8.25 points over BAGEL. Both single routes yield smaller gains in caption detail retention and improve color binding on this diagnostic. Composition further raises detail retention above either route and retains nearly all of G-only's color-binding gain. It therefore brings improvements in both caption detail retention and visual detail realization into a single model, addressing the two failures examined in Figure~\ref{fig:motivation}.

\subsection{Cross-task consistency and synergy}\label{sec:consistency}

\begin{table}[!htbp]
\centering
\caption{\textbf{Cross-task consistency and synergy on BAGEL.}}\label{tab:consistency}
\tableformat
\setlength{\tabcolsep}{10pt}
\begin{tabular}{lrr|rrrrr}
\toprule
\rowcolor{tablehead}
 & \multicolumn{2}{c|}{\textbf{RealUnify}} & \multicolumn{5}{c}{\textbf{Unison}}\\
\rowcolor{tablehead}
\textbf{Method} & GEU & UEG & IC & UGG & GGU & ME & Mean\\
\midrule
BAGEL & 39.00 & 22.00 & 80.51 & 72.68 & 32.65 & 35.41 & 55.31\\
\rowcolor{oursbg}
\textbf{\method} & 40.00 & 23.67 & 79.57 & 71.33 & 33.20 & 57.40 & 60.37\\
\bottomrule
\end{tabular}
\end{table}

Separate task scores do not establish how well understanding and generation work together within a task. We therefore evaluate their interaction on RealUnify and Unison (Table~\ref{tab:consistency}). RealUnify measures generation-enhanced understanding (GEU, 400 examples) and understanding-enhanced generation (UEG, 600 examples). Both use the Direct protocol, with Qwen2.5-VL-72B-Instruct judging UEG. Unison covers internal consistency (IC, 556 examples), understanding-guided generation (UGG, 543), generation-guided understanding (GGU, 542), and mutual enhancement (ME, 264), and summarizes these capabilities using an equally weighted mean across the four tasks.

\method\ improves both RealUnify directions and raises the mean Unison score from 55.31 to 60.37. The largest change is in mutual enhancement. Part of this gain comes from better completion of the required response format: examples without a parseable mismatch list fall from 234 to 25 out of 264, and the evaluator assigns zero understanding credit to such responses. The ME increase therefore reflects improved format compliance as well as task performance. The gains in GEU, UEG, and GGU show benefits across several interaction patterns, while the lower IC and UGG scores indicate that adaptation does not improve every form of cross-task consistency.

\Needspace{25\baselineskip}
\subsection{Transfer across architectures}
\noindent
\begin{minipage}[t]{0.45\linewidth}
\vspace{0pt}
\raggedright
We test whether the target construction remains useful when the backbone's visual prediction objective changes. Table~\ref{tab:transfer} applies the same two routes and composition rule to the autoregressive Janus-Pro and continuous-generation Show-o2, retaining each model's native objective. Both backbones improve on MMMU, MathVista, MMStar, and overall WISE. Janus-Pro gains 5.93 points on MMStar, while Show-o2 gains 0.98 points on WISE. This agreement across architectures supports constructing detail-targeted supervision at the image--text level, where it can be used by different native generators.
\end{minipage}\hfill
\begin{minipage}[t]{0.52\linewidth}
\vspace{0pt}
\centering
\captionsetup{hypcap=false,justification=raggedright,singlelinecheck=false}
\captionof{table}{\textbf{Transfer across backbones.} Base denotes the TorchUMM reference, and Ours denotes \method.}\label{tab:transfer}
\tableformat
\footnotesize
\setlength{\tabcolsep}{2.2pt}
\begin{tabularx}{\linewidth}{Lrr|rr}
\toprule
\rowcolor{tablehead}
 & \multicolumn{2}{c|}{\textbf{Janus-Pro}} & \multicolumn{2}{c}{\textbf{Show-o2}}\\
\rowcolor{tablehead}
\textbf{Benchmark} & Base & \cellcolor{oursbg}Ours & Base & \cellcolor{oursbg}Ours\\
\midrule
\multicolumn{5}{l}{\textit{Understanding}}\\
MMMU & 40.70 & 41.00 & 47.90 & 48.00\\
MMBench & 67.64 & 68.05 & 77.25 & 77.25\\
MathVista & 42.80 & 44.20 & 51.50 & 51.60\\
MMStar & 42.00 & 47.93 & 54.53 & 55.60\\
\midrule
\multicolumn{5}{l}{\textit{Generation (WISE)}}\\
Culture & 36.16 & 35.05 & 36.41 & 36.78\\
Time & 38.53 & 39.34 & 34.97 & 37.46\\
Space & 47.89 & 47.03 & 45.19 & 47.74\\
Biology & 36.05 & 41.30 & 34.55 & 37.50\\
Physics & 47.45 & 47.75 & 36.90 & 35.30\\
Chemistry & 24.85 & 26.75 & 23.90 & 23.35\\
Overall & 38.11 & 38.43 & 35.95 & 36.93\\
\bottomrule
\end{tabularx}
\end{minipage}

\subsection{Ablation of the two training routes}\label{sec:ablations}
To separate the contributions of the two objectives, Table~\ref{tab:routes} compares generation-only adaptation (G-only), understanding-only adaptation (U-only), and their composition under common evaluation settings, using the pretrained reference reported by TorchUMM. G-only improves both generation benchmarks and all three reported understanding benchmarks. U-only obtains the strongest MMStar result among the adapted variants, but its WISE and UEval scores are slightly below the pretrained reference. Its color-binding improvement in Figure~\ref{fig:detail_diagnostics} therefore does not extend to an overall gain on these generation benchmarks. Composition improves all three understanding scores over G-only while preserving nearly all of its WISE gain, scoring 43.48 compared with 43.52, and obtains the highest UEval score among the variants. Its MMMU score rises from 51.90 to 53.30, a gain of 1.40 percentage points over the pretrained reference. This pattern supports composing independently learned updates to retain strong generation performance while improving understanding, with the benefits depending on the evaluated capability.

\begin{table}[htbp]
\centering
\begin{minipage}{0.88\linewidth}
\caption{\textbf{Ablation of the two training routes on BAGEL.} Bold and underline mark the best and second-best available results.}\label{tab:routes}
\tableformat
\setlength{\tabcolsep}{1.4pt}
\begin{tabular*}{\linewidth}{@{\extracolsep{\fill}}lrrrrr@{}}
\toprule
\rowcolor{tablehead}
 & \multicolumn{3}{c}{\textbf{Understanding}} & \multicolumn{2}{c}{\textbf{Generation}}\\
\rowcolor{tablehead}
\multirow{-2}{*}{\textbf{Variant}} & MMMU & MMStar & MathVista & WISE & UEval\\
\midrule
BAGEL & 51.90 & 63.20 & 71.60 & 39.89 & 30.90\\
G-only & \underline{53.00} & 68.73 & 73.10 & \textbf{43.52} & \underline{31.33}\\
U-only & \underline{53.00} & \textbf{70.40} & \underline{73.20} & 39.67 & 30.48\\
\rowcolor{oursbg}
\textbf{\method} & \textbf{53.30} & \underline{70.13} & \textbf{73.80} & \underline{43.48} & \textbf{31.60}\\
\bottomrule
\end{tabular*}
\end{minipage}
\end{table}

\noindent
\begin{minipage}[t]{0.51\linewidth}
\vspace{0pt}
\raggedright
\paragraph{Composition weight.}
Figure~\ref{fig:composition_weight} varies the understanding-update weight $\beta$ while fixing $\alpha=1.0$. Compared with generation-only adaptation ($\beta=0$), $\beta=0.02$ raises MMBench-Dev accuracy from 79.2\% to 82.0\% while maintaining the GenEval score at 75.0. Larger weights reduce generation performance. At $\beta=0.2$, understanding reaches 82.6\%, but GenEval falls to 64.0. The selected weight therefore improves understanding while preserving the generation score on the development set.
\end{minipage}\hfill
\begin{minipage}[t]{0.45\linewidth}
\vspace{0pt}
\centering
\includegraphics[width=\linewidth]{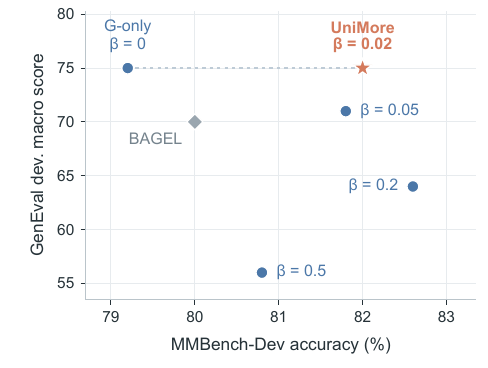}
\captionsetup{hypcap=false,justification=raggedright,singlelinecheck=false}
\captionof{figure}{\textbf{Composition weight on BAGEL development data.} The star marks the selected $\beta=0.02$.}\label{fig:composition_weight}
\end{minipage}

\subsection{Effect of supervision targets}\label{sec:supervision_targets}
The route ablation leaves open whether the gains depend on how the paired data are organized into supervision or merely on additional training. We therefore vary the prediction and comparison targets within each route while keeping its source pool and optimizer-update budget fixed (Table~\ref{tab:supervision_targets}).

\begin{table}[!htbp]
\centering
\caption{\textbf{Supervision-target comparisons on BAGEL.} (a) Visual prediction targets. (b) Caption supervision, where ``+'' adds ranking to positive-caption CE.}\label{tab:supervision_targets}
\tableformat
\setlength{\tabcolsep}{4pt}
\begin{minipage}[t]{0.30\linewidth}
\centering
\textbf{(a) Generation targets}\par\smallskip
\begin{tabularx}{\linewidth}{Lr}
\toprule
\rowcolor{tablehead}
\textbf{Supervision} & \textbf{WISE}\\
\midrule
BAGEL & 39.89\\
\midrule
Caption $\to$ Image & 40.18\\
Phrase $\to$ Image & 42.00\\
Phrase $\to$ Crop & 43.52\\
\midrule
\rowcolor{oursbg}
\textbf{\method} & 43.48\\
\bottomrule
\end{tabularx}
\end{minipage}\hfill
\begin{minipage}[t]{0.66\linewidth}
\centering
\textbf{(b) Caption supervision}\par\smallskip
\setlength{\tabcolsep}{3pt}
\begin{tabularx}{\linewidth}{Lrrrr}
\toprule
\rowcolor{tablehead}
\textbf{Supervision} & \textbf{MMMU} & \textbf{MMStar} & \textbf{MathVista} & \textbf{Avg.}\\
\midrule
BAGEL & 51.90 & 63.20 & 71.60 & 62.23\\
\midrule
SFT  & 52.60 & 68.27 & 73.10 & 64.66\\
+ Deletion & 52.20 & 70.20 & 73.20 & 65.20\\
\mbox{+ Deletion + corruption} & 53.00 & 70.40 & 73.20 & 65.53\\
\midrule
\rowcolor{oursbg}
\textbf{\method} & 53.30 & 70.13 & 73.80 & 65.74\\
\bottomrule
\end{tabularx}
\end{minipage}
\end{table}

\paragraph{Generation targets.}
We compare full-caption prediction of the full image, grounded-phrase prediction of the full image, and grounded-phrase prediction of the corresponding crop. All variants use 2,460 source records, 1,000 optimizer updates, four global packed microbatches per update, and the same WISE evaluation settings. The full-image variants use two GPUs with two accumulation steps, while the phrase--crop variant distributes its training across four GPUs.

WISE increases from 40.18 with full-caption supervision to 42.00 with phrase conditioning, then to 43.52 when the visual target is also restricted to the grounded crop. The additional gain from changing the visual target supports matching the scope of the condition and prediction. A crop concentrates the visual training signal on the content named by the phrase, providing a direct learning target for the requested detail. This progression is consistent with the design rationale.

\paragraph{Caption supervision.}
We test whether detail-aware caption comparisons improve on positive-caption fitting.
The three variants use positive-caption cross entropy alone, ranking against one detail-deleted caption and two random captions, or ranking against one deleted, one corrupted, and one random caption. Avg. in Table~\ref{tab:supervision_targets} is the unweighted mean of MMMU, MMStar, and MathVista.

The full alternative set yields the highest MMMU and MMStar scores among these variants. Adding corruption in place of a random alternative raises MMMU from 52.20 to 53.00 without reducing MMStar or MathVista. This comparison suggests that incorrect details provide a useful distinction beyond missing details and unrelated descriptions of other scenes.

\subsection{Detail realization and recognition}\label{sec:detail_results}

\begin{figure}[!htbp]
\centering
\includegraphics[width=\textwidth]{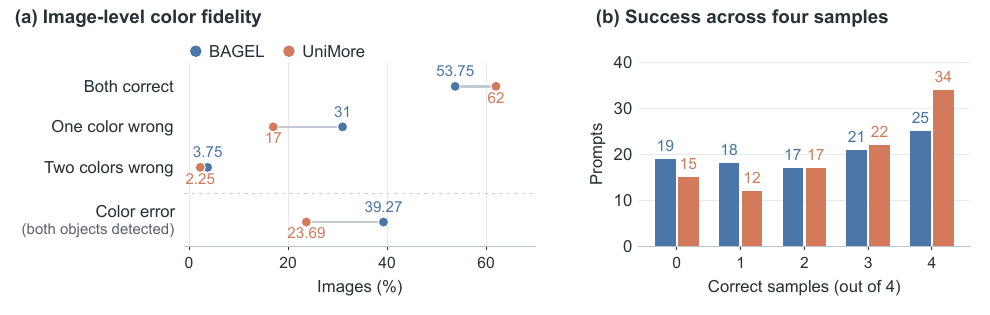}
\caption{\textbf{Color fidelity and repeated sampling.} (a) Color outcomes, with the bottom row conditioned on both objects being detected. (b) Prompts grouped by correct samples out of four.}\label{fig:color_fidelity}
\end{figure}

\paragraph{Color fidelity and repeated sampling.}
To examine the color-binding improvement in Figure~\ref{fig:detail_diagnostics}, we further analyze object detection and success across repeated samples for pretrained BAGEL and \method. The diagnostic uses 100 GenEval color-attribution prompts, each specifying two distinct object types with different colors, and four images per prompt. Both models use $512\times512$ generation, 50 sampling steps, seed 42, text guidance 4, and image guidance 1. Object and color labels follow the same GenEval detector and color classifier.
Each of the first three outcomes in Figure~\ref{fig:color_fidelity}a requires both objects to be detected, while its percentage is computed over all generated images. The separate bottom row instead conditions on both objects being detected, for which color error decreases from 39.27\% to 23.69\%.

\method\ improves joint object-and-color correctness from 215 to 248 of the 400 images, a gain of 8.25 percentage points (Figure~\ref{fig:color_fidelity}a).  Color accuracy also improves by 14.10 points on the 305 matched pairs with complete detections for both models. This matched comparison supports improved attribute realization when both models generate detectable target objects.
Figure~\ref{fig:color_fidelity}b groups prompts by their number of correct images among four, showing that the gain extends to repeated success on the same prompt. The number of prompts with four correct images increases from 25 to 34. Thus, the improvement includes more prompts for which all four samples realize the requested details.

\Needspace{16\baselineskip}
\begin{wrapfigure}{r}{0.53\textwidth}
\centering
\captionsetup{justification=raggedright,singlelinecheck=false}
\includegraphics[width=\linewidth]{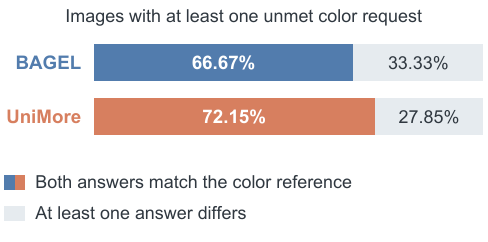}
\caption{\textbf{Describing colors after generation failure.} Among images with at least one unmet color request, colored segments indicate agreement with the automatic reference for both objects.}\label{fig:self_color}
\end{wrapfigure}

\paragraph{Color recognition on self-generated images.}
We examine whether failure to generate a requested color also entails failure to describe the color that appears in the generated image. BAGEL and \method\ each answer questions about their own color-attribution images described above. We ask about each target object separately. Each question uses a fresh image context without the original prompt or requested color. Decoding is greedy without thinking, with a 500-token limit. Answers use 11 basic colors or indicate that the object is missing, unclear, multicolored, or transparent.

Qwen2.5-VL-32B-Instruct annotates object colors from the image and object question, independently of the requested colors and model answers. We evaluate images with two definite reference colors, counting incorrect or unparseable model answers as errors. Generation fulfillment requires both reference colors to match the request, while recognition agreement requires both model answers to match the reference. Both measures use these automatic reference annotations.

We focus on images where at least one reference color differs from the requested color. Each full bar in Figure~\ref{fig:self_color} represents this failure subset, and its colored segment shows images where both model answers still match the reference. BAGEL does so on 66.67\% of these images, and \method\ on 72.15\%. Agreement also remains frequent for individual incorrectly colored objects, at 71.43\% and 75.82\%, respectively. Thus, a model can describe the color that appears in its image even when it fails to produce the requested color. This separation helps explain why supervision for visual realization remains useful alongside the model's ability to answer questions about visual details.

\begin{figure}[!tbp]
\centering
\includegraphics[width=\textwidth]{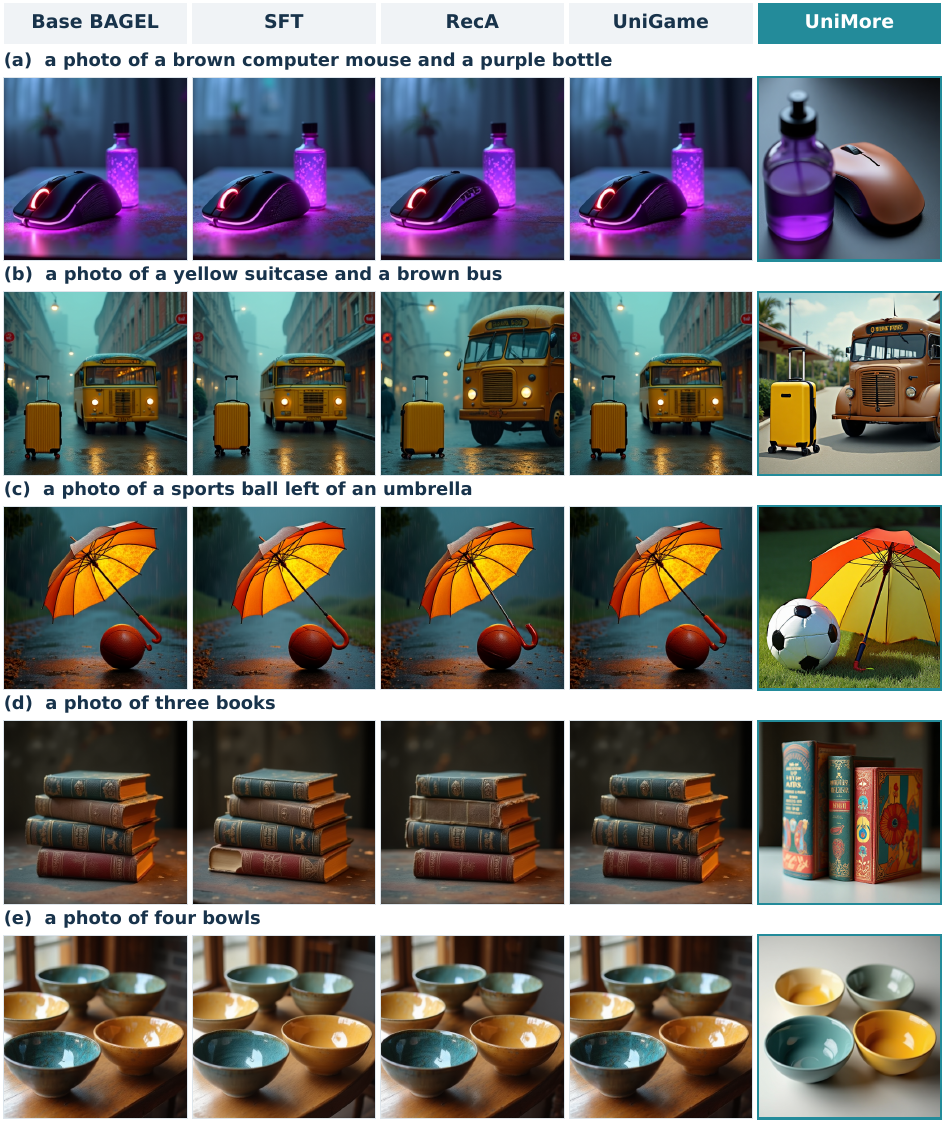}
\caption{\textbf{Qualitative comparisons on GenEval}, covering colors (a, b), spatial relations (c), and object counts (d, e).}\label{fig:qualitative}
\end{figure}

\paragraph{Qualitative generation results.}
We compare generated images to show how the quantitative gains appear in concrete attribute, relation, and count requirements (Figure~\ref{fig:qualitative}). Under the same prompts and sampling settings, \method\ assigns the requested colors to a brown mouse and purple bottle, and to a yellow suitcase and brown bus. It also places a ball to the left of an umbrella and generates the requested three books and four bowls, where the reference outputs contain an extra object. These successful examples show that the improved details can affect both an object's appearance and the spatial arrangement or exact number of objects depicted in a scene.

\subsection{Adaptation cost}\label{sec:cost}
We report training time and adapter size to characterize the cost of applying \method\ to additional backbones. Janus-Pro and Show-o2 each use one H100 GPU per route. Their two routes require approximately 0.595 and 0.356 GPU-hours in total, respectively. These measurements cover the training loop and adapter writing, excluding model loading, target construction, evaluation, and hyperparameter search.

Each route trains 37.48 million LoRA parameters for Janus-Pro and 40.37 million for Show-o2, with saved adapters of approximately 71.5 and 77.0 MiB. Thus, the measured adaptation requires less than one GPU-hour on each backbone and stores only compact parameter updates. Static composition incorporates these updates into the original model without adding deployed parameters.

%% file: section/conclusion.tex
\section{Conclusion}\label{sec:conclusion}

We introduced \method\ to learn more from image--text pairs by organizing their specific content into visual prediction and textual comparison tasks. The method pairs grounded phrase-to-crop prediction with full-image caption discrimination, then composes the independently adapted updates into a single model. Evaluation across three backbones shows that this target construction works with different native visual objectives and improves understanding and generation benchmarks. Paired diagnostics identify improved compositional color fidelity, while the training studies examine how supervision and composition affect downstream performance. These results support supervision construction as a practical approach to adapting both prediction directions of unified multimodal models using existing image--text pairs.

%% file: section/declarations.tex
\section*{Code and Data}

\paragraph{Data availability}
The UniMore dataset is available at \url{https://huggingface.co/datasets/HuggingSelf/UniMore}.

\paragraph{Code and model availability}
The code is available in TorchUMM at \url{https://github.com/AIFrontierLab/TorchUMM/tree/main/src/umm/post_training/UniMore}. The model checkpoints are available at \url{https://huggingface.co/HuggingSelf/UniMore}.

%% file: refs.bib
@inproceedings{nextgpt,
  title = {{NExT-GPT: Any-to-Any Multimodal LLM}},
  author = {Wu, Shengqiong and Fei, Hao and Qu, Leigang and Ji, Wei and Chua, Tat-Seng},
  booktitle = {Proceedings of the 41st International Conference on Machine Learning},
  pages = {53366--53397},
  volume = {235},
  series = {Proceedings of Machine Learning Research},
  year = {2024},
  publisher = {PMLR},
  url = {https://proceedings.mlr.press/v235/wu24e.html}
}

@inproceedings{vitron,
  title = {{VITRON: A Unified Pixel-level Vision LLM for Understanding, Generating, Segmenting, Editing}},
  author = {Fei, Hao and Wu, Shengqiong and Zhang, Hanwang and Chua, Tat-Seng and Yan, Shuicheng},
  booktitle = {Advances in Neural Information Processing Systems},
  volume = {37},
  year = {2024},
  url = {https://proceedings.neurips.cc/paper_files/paper/2024/hash/68bad5506f0f9eea7ae75f01ae00d5e2-Abstract-Conference.html}
}

@inproceedings{xfusion,
  title = {{X-Fusion: Introducing New Modality to Frozen Large Language Models}},
  author = {Mo, Sicheng and Nguyen, Thao and Huang, Xun and Iyer, Siddharth Srinivasan and Li, Yijun and Liu, Yuchen and Tandon, Abhishek and Shechtman, Eli and Singh, Krishna Kumar and Lee, Yong Jae and Zhou, Bolei and Li, Yuheng},
  booktitle = {Proceedings of the IEEE/CVF International Conference on Computer Vision},
  pages = {228--238},
  year = {2025},
  url = {https://openaccess.thecvf.com/content/ICCV2025/html/Mo_X-Fusion_Introducing_New_Modality_to_Frozen_Large_Language_Models_ICCV_2025_paper.html}
}

@inproceedings{tar,
  title = {{Vision as a Dialect: Unifying Visual Understanding and Generation via Text-Aligned Representations}},
  author = {Han, Jiaming and Chen, Hao and Zhao, Yang and Wang, Hanyu and Zhao, Qi and Yang, Ziyan and He, Hao and Yue, Xiangyu and Jiang, Lu},
  booktitle = {Advances in Neural Information Processing Systems},
  volume = {38},
  year = {2025},
  url = {https://papers.nips.cc/paper_files/paper/2025/hash/e87b1e06be8c3594c810e8991e77ea40-Abstract-Conference.html}
}

@misc{unirl,
  title = {{UniRL: Self-Improving Unified Multimodal Models via Supervised and Reinforcement Learning}},
  author = {Mao, Weijia and Yang, Zhenheng and Shou, Mike Zheng},
  year = {2025},
  eprint = {2505.23380},
  archivePrefix = {arXiv},
  url = {https://arxiv.org/abs/2505.23380}
}

@inproceedings{adloop,
  title = {{Synergizing Understanding and Generation with Interleaved Analyzing-Drafting Thinking}},
  author = {Wu, Shengqiong and Li, Bobo and Wang, Xinkai and Li, Xiangtai and Cui, Lei and Wei, Furu and Yan, Shuicheng and Fei, Hao and Chua, Tat-Seng},
  booktitle = {International Conference on Learning Representations},
  year = {2026},
  url = {https://openreview.net/forum?id=GtqmPJf00A}
}

@misc{native_synergy,
  title = {{Uncovering Understanding-Generation Synergy in Native Unified Multimodal Models: From Representation, Task to System}},
  author = {Wu, Penghao and Diao, Haiwen and Fan, Weichen and Lu, Lewei and Lin, Dahua and Liu, Ziwei},
  year = {2026},
  eprint = {2609.01607},
  archivePrefix = {arXiv},
  url = {https://arxiv.org/abs/2609.01607}
}

@inproceedings{glamm,
  title = {{GLaMM: Pixel Grounding Large Multimodal Model}},
  author = {Rasheed, Hanoona and Maaz, Muhammad and Shaji, Sahal and Shaker, Abdelrahman and Khan, Salman and Cholakkal, Hisham and Anwer, Rao M. and Xing, Eric and Yang, Ming-Hsuan and Khan, Fahad S.},
  booktitle = {Proceedings of the IEEE/CVF Conference on Computer Vision and Pattern Recognition},
  pages = {13009--13018},
  year = {2024},
  url = {https://openaccess.thecvf.com/content/CVPR2024/html/Rasheed_GLaMM_Pixel_Grounding_Large_Multimodal_Model_CVPR_2024_paper.html}
}

@inproceedings{generateu,
  title = {{Generative Region-Language Pretraining for Open-Ended Object Detection}},
  author = {Lin, Chuang and Jiang, Yi and Qu, Lizhen and Yuan, Zehuan and Cai, Jianfei},
  booktitle = {Proceedings of the IEEE/CVF Conference on Computer Vision and Pattern Recognition},
  pages = {13958--13968},
  year = {2024},
  url = {https://openaccess.thecvf.com/content/CVPR2024/html/Lin_Generative_Region-Language_Pretraining_for_Open-Ended_Object_Detection_CVPR_2024_paper.html}
}

@inproceedings{clipreward,
  title = {{Fine-grained Image Captioning with CLIP Reward}},
  author = {Cho, Jaemin and Yoon, Seunghyun and Kale, Ajinkya and Dernoncourt, Franck and Bui, Trung and Bansal, Mohit},
  booktitle = {Findings of the Association for Computational Linguistics: NAACL 2022},
  pages = {517--527},
  year = {2022},
  doi = {10.18653/v1/2022.findings-naacl.39},
  url = {https://aclanthology.org/2022.findings-naacl.39/}
}

@misc{blip3o,
  title = {{BLIP3-o: A Family of Fully Open Unified Multimodal Models-Architecture, Training and Dataset}},
  author = {Jiuhai Chen and Zhiyang Xu and Xichen Pan and Yushi Hu and Can Qin and Tom Goldstein and Lifu Huang and Tianyi Zhou and Saining Xie and Silvio Savarese and Le Xue and Caiming Xiong and Ran Xu},
  year = {2025},
  eprint = {2505.09568},
  archivePrefix = {arXiv},
  url = {https://arxiv.org/abs/2505.09568}
}

@inproceedings{tokenflow,
  title = {{TokenFlow: Unified Image Tokenizer for Multimodal Understanding and Generation}},
  author = {Qu, Liao and Zhang, Huichao and Liu, Yiheng and Wang, Xu and Jiang, Yi and Gao, Yiming and Ye, Hu and Du, Daniel K. and Yuan, Zehuan and Wu, Xinglong},
  booktitle = {Proceedings of the IEEE/CVF Conference on Computer Vision and Pattern Recognition},
  pages = {2545--2555},
  year = {2025},
  url = {https://openaccess.thecvf.com/content/CVPR2025/html/Qu_TokenFlow_Unified_Image_Tokenizer_for_Multimodal_Understanding_and_Generation_CVPR_2025_paper.html}
}

@misc{deepgen,
  title = {{DeepGen 1.0: A Lightweight Unified Multimodal Model for Advancing Image Generation and Editing}},
  author = {Dianyi Wang and Ruihang Li and Feng Han and Chaofan Ma and Wei Song and Siyuan Wang and Yibin Wang and Yi Xin and Hongjian Liu and Zhixiong Zhang and Shengyuan Ding and Tianhang Wang and Zhenglin Cheng and Tao Lin and Cheng Jin and Kaicheng Yu and Jingjing Chen and Wenjie Wang and Zhongyu Wei and Jiaqi Wang},
  year = {2026},
  eprint = {2602.12205},
  archivePrefix = {arXiv},
  url = {https://arxiv.org/abs/2602.12205}
}

@inproceedings{pcgrad,
  author = {Yu, Tianhe and Kumar, Saurabh and Gupta, Abhishek and Levine, Sergey and Hausman, Karol and Finn, Chelsea},
  title = {Gradient Surgery for Multi-Task Learning},
  booktitle = {Advances in Neural Information Processing Systems},
  volume = {33},
  pages = {5824--5836},
  publisher = {Curran Associates, Inc.},
  year = {2020},
  url = {https://proceedings.neurips.cc/paper_files/paper/2020/file/3fe78a8acf5fda99de95303940a2420c-Paper.pdf},
}

@misc{bagel,
      title={{Emerging Properties in Unified Multimodal Pretraining}},
      author={Chaorui Deng and Deyao Zhu and Kunchang Li and Chenhui Gou and Feng Li and Zeyu Wang and Shu Zhong and Weihao Yu and Xiaonan Nie and Ziang Song and Guang Shi and Haoqi Fan},
      year={2025},
      eprint={2505.14683},
      archivePrefix={arXiv},
      primaryClass={cs.CV},
      url={https://arxiv.org/abs/2505.14683},
}

@misc{januspro,
      title={{Janus-Pro: Unified Multimodal Understanding and Generation with Data and Model Scaling}},
      author={Xiaokang Chen and Zhiyu Wu and Xingchao Liu and Zizheng Pan and Wen Liu and Zhenda Xie and Xingkai Yu and Chong Ruan},
      year={2025},
      eprint={2501.17811},
      archivePrefix={arXiv},
      primaryClass={cs.AI},
      url={https://arxiv.org/abs/2501.17811},
}

@inproceedings{showo2,
  author = {Xie, Jinheng and Yang, Zhenheng and Shou, Mike Zheng},
  title = {{Show-o2: Improved Native Unified Multimodal Models}},
  booktitle = {Advances in Neural Information Processing Systems},
  volume = {38},
  pages = {47490--47518},
  publisher = {Curran Associates, Inc.},
  year = {2025},
  doi = {10.52202/085713-1584},
  url = {https://proceedings.neurips.cc/paper_files/paper/2025/hash/43e8fd8b9581faa71a6a61602bc28435-Abstract-Conference.html}
}

@misc{fgclip,
      title={{FG-CLIP: Fine-Grained Visual and Textual Alignment}},
      author={Chunyu Xie and Bin Wang and Fanjing Kong and Jincheng Li and Dawei Liang and Gengshen Zhang and Dawei Leng and Yuhui Yin},
      year={2025},
      eprint={2505.05071},
      archivePrefix={arXiv},
      primaryClass={cs.CV},
      url={https://arxiv.org/abs/2505.05071},
}

@misc{uno,
      title={{Steering Visual Generation in Unified Multimodal Models with Understanding Supervision}},
      author={Zeyu Liu and Zanlin Ni and Yang Yue and Cheng Da and Huan Yang and Di Zhang and Kun Gai and Gao Huang},
      year={2026},
      eprint={2605.05781},
      archivePrefix={arXiv},
      primaryClass={cs.CV},
      url={https://arxiv.org/abs/2605.05781},
}

@misc{srum,
      title={{SRUM: Fine-Grained Self-Rewarding for Unified Multimodal Models}},
      author={Weiyang Jin and Yuwei Niu and Jiaqi Liao and Chengqi Duan and Aoxue Li and Shenghua Gao and Xihui Liu},
      year={2026},
      eprint={2510.12784},
      archivePrefix={arXiv},
      primaryClass={cs.CV},
      url={https://arxiv.org/abs/2510.12784},
}

@misc{corl,
      title={{Co-Reinforcement Learning for Unified Multimodal Understanding and Generation}},
      author={Jingjing Jiang and Chongjie Si and Jun Luo and Hanwang Zhang and Chao Ma},
      year={2025},
      eprint={2505.17534},
      archivePrefix={arXiv},
      primaryClass={cs.CV},
      url={https://arxiv.org/abs/2505.17534},
}

@inproceedings{reca,
  author = {Xie, Ji and Darrell, Trevor and Zettlemoyer, Luke and Wang, Xudong},
  title = {{Reconstruction Alignment Improves Unified Multimodal Models}},
  booktitle = {International Conference on Learning Representations},
  pages = {120095--120137},
  year = {2026},
  url = {https://proceedings.iclr.cc/paper_files/paper/2026/hash/c34ddd05eb089991f06f3c5dc36836e0-Abstract-Conference.html}
}

@inproceedings{unigame,
  author = {Su, Zhaolong and Lu, Wang and Chen, Hao and Li, Sharon and Wang, Jindong},
  title = {{UniGame: Turning a Unified Multimodal Model Into Its Own Adversary}},
  booktitle = {Proceedings of the IEEE/CVF Conference on Computer Vision and Pattern Recognition},
  pages = {37632--37641},
  year = {2026},
  url = {https://openaccess.thecvf.com/content/CVPR2026/html/Su_UniGame_Turning_a_Unified_Multimodal_Model_Into_Its_Own_Adversary_CVPR_2026_paper.html}
}

@misc{gvu,
      title={{Learning to Generate via Understanding: Understanding-Driven Intrinsic Rewarding for Unified Multimodal Models}},
      author={Jiadong Pan and Liang Li and Yuxin Peng and Yu-Ming Tang and Shuohuan Wang and Yu Sun and Hua Wu and Qingming Huang and Haifeng Wang},
      year={2026},
      eprint={2603.06043},
      archivePrefix={arXiv},
      primaryClass={cs.CV},
      url={https://arxiv.org/abs/2603.06043},
}

@misc{lora,
      title={{LoRA: Low-Rank Adaptation of Large Language Models}},
      author={Edward J. Hu and Yelong Shen and Phillip Wallis and Zeyuan Allen-Zhu and Yuanzhi Li and Shean Wang and Lu Wang and Weizhu Chen},
      year={2021},
      eprint={2106.09685},
      archivePrefix={arXiv},
      primaryClass={cs.CL},
      url={https://arxiv.org/abs/2106.09685},
}

@misc{groundingdino,
      title={{Grounding DINO: Marrying DINO with Grounded Pre-Training for Open-Set Object Detection}},
      author={Shilong Liu and Zhaoyang Zeng and Tianhe Ren and Feng Li and Hao Zhang and Jie Yang and Qing Jiang and Chunyuan Li and Jianwei Yang and Hang Su and Jun Zhu and Lei Zhang},
      year={2024},
      eprint={2303.05499},
      archivePrefix={arXiv},
      primaryClass={cs.CV},
      url={https://arxiv.org/abs/2303.05499},
}

@misc{clip,
      title={{Learning Transferable Visual Models From Natural Language Supervision}},
      author={Alec Radford and Jong Wook Kim and Chris Hallacy and Aditya Ramesh and Gabriel Goh and Sandhini Agarwal and Girish Sastry and Amanda Askell and Pamela Mishkin and Jack Clark and Gretchen Krueger and Ilya Sutskever},
      year={2021},
      eprint={2103.00020},
      archivePrefix={arXiv},
      primaryClass={cs.CV},
      url={https://arxiv.org/abs/2103.00020},
}

@misc{geneval,
      title={{GenEval: An Object-Focused Framework for Evaluating Text-to-Image Alignment}},
      author={Dhruba Ghosh and Hanna Hajishirzi and Ludwig Schmidt},
      year={2023},
      eprint={2310.11513},
      archivePrefix={arXiv},
      primaryClass={cs.CV},
      url={https://arxiv.org/abs/2310.11513},
}

@misc{mmmu,
      title={{MMMU: A Massive Multi-discipline Multimodal Understanding and Reasoning Benchmark for Expert AGI}},
      author={Xiang Yue and Yuansheng Ni and Kai Zhang and Tianyu Zheng and Ruoqi Liu and Ge Zhang and Samuel Stevens and Dongfu Jiang and Weiming Ren and Yuxuan Sun and Cong Wei and Botao Yu and Ruibin Yuan and Renliang Sun and Ming Yin and Boyuan Zheng and Zhenzhu Yang and Yibo Liu and Wenhao Huang and Huan Sun and Yu Su and Wenhu Chen},
      year={2024},
      eprint={2311.16502},
      archivePrefix={arXiv},
      primaryClass={cs.CL},
      url={https://arxiv.org/abs/2311.16502},
}

@misc{mmbench,
      title={{MMBench: Is Your Multi-modal Model an All-around Player?}},
      author={Yuan Liu and Haodong Duan and Yuanhan Zhang and Bo Li and Songyang Zhang and Wangbo Zhao and Yike Yuan and Jiaqi Wang and Conghui He and Ziwei Liu and Kai Chen and Dahua Lin},
      year={2024},
      eprint={2307.06281},
      archivePrefix={arXiv},
      primaryClass={cs.CV},
      url={https://arxiv.org/abs/2307.06281},
}

@misc{mathvista,
      title={{MathVista: Evaluating Mathematical Reasoning of Foundation Models in Visual Contexts}},
      author={Pan Lu and Hritik Bansal and Tony Xia and Jiacheng Liu and Chunyuan Li and Hannaneh Hajishirzi and Hao Cheng and Kai-Wei Chang and Michel Galley and Jianfeng Gao},
      year={2024},
      eprint={2310.02255},
      archivePrefix={arXiv},
      primaryClass={cs.CV},
      url={https://arxiv.org/abs/2310.02255},
}

@misc{taskarith,
      title={{Editing Models with Task Arithmetic}},
      author={Gabriel Ilharco and Marco Tulio Ribeiro and Mitchell Wortsman and Suchin Gururangan and Ludwig Schmidt and Hannaneh Hajishirzi and Ali Farhadi},
      year={2023},
      eprint={2212.04089},
      archivePrefix={arXiv},
      primaryClass={cs.LG},
      url={https://arxiv.org/abs/2212.04089},
}

@misc{tokencompose,
      title={{TokenCompose: Text-to-Image Diffusion with Token-level Supervision}},
      author={Zirui Wang and Zhizhou Sha and Zheng Ding and Yilin Wang and Zhuowen Tu},
      year={2024},
      eprint={2312.03626},
      archivePrefix={arXiv},
      primaryClass={cs.CV},
      url={https://arxiv.org/abs/2312.03626},
}

@misc{docci,
      title={{DOCCI: Descriptions of Connected and Contrasting Images}},
      author={Yasumasa Onoe and Sunayana Rane and Zachary Berger and Yonatan Bitton and Jaemin Cho and Roopal Garg and Alexander Ku and Zarana Parekh and Jordi Pont-Tuset and Garrett Tanzer and Su Wang and Jason Baldridge},
      year={2024},
      eprint={2404.19753},
      archivePrefix={arXiv},
      primaryClass={cs.CV},
      url={https://arxiv.org/abs/2404.19753},
}

@misc{dsg,
      title={{Davidsonian Scene Graph: Improving Reliability in Fine-grained Evaluation for Text-to-Image Generation}},
      author={Jaemin Cho and Yushi Hu and Roopal Garg and Peter Anderson and Ranjay Krishna and Jason Baldridge and Mohit Bansal and Jordi Pont-Tuset and Su Wang},
      year={2024},
      eprint={2310.18235},
      archivePrefix={arXiv},
      primaryClass={cs.CV},
      url={https://arxiv.org/abs/2310.18235},
}

@misc{mmstar,
      title={{Are We on the Right Way for Evaluating Large Vision-Language Models?}},
      author={Lin Chen and Jinsong Li and Xiaoyi Dong and Pan Zhang and Yuhang Zang and Zehui Chen and Haodong Duan and Jiaqi Wang and Yu Qiao and Dahua Lin and Feng Zhao},
      year={2024},
      eprint={2403.20330},
      archivePrefix={arXiv},
      primaryClass={cs.CV},
      url={https://arxiv.org/abs/2403.20330},
}

@misc{wise,
      title={{WISE: A World Knowledge-Informed Semantic Evaluation for Text-to-Image Generation}},
      author={Yuwei Niu and Munan Ning and Mengren Zheng and Weiyang Jin and Bin Lin and Peng Jin and Jiaqi Liao and Chaoran Feng and Fanqing Meng and Kunpeng Ning and Bin Zhu and Li Yuan},
      year={2026},
      eprint={2503.07265},
      archivePrefix={arXiv},
      primaryClass={cs.CV},
      url={https://arxiv.org/abs/2503.07265},
}

@misc{realunify,
      title={{RealUnify: Do Unified Models Truly Benefit from Unification? A Comprehensive Benchmark}},
      author={Yang Shi and Yuhao Dong and Yue Ding and Yuran Wang and Xuanyu Zhu and Sheng Zhou and Wenting Liu and Haochen Tian and Rundong Wang and Huanqian Wang and Zuyan Liu and Bohan Zeng and Ruizhe Chen and Qixun Wang and Zhuoran Zhang and Xinlong Chen and Chengzhuo Tong and Bozhou Li and Qiang Liu and Haotian Wang and Wenjing Yang and Yuanxing Zhang and Pengfei Wan and Yi-Fan Zhang and Ziwei Liu},
      year={2026},
      eprint={2509.24897},
      archivePrefix={arXiv},
      primaryClass={cs.AI},
      url={https://arxiv.org/abs/2509.24897},
}

@misc{ueval,
      title={{UEval: A Benchmark for Unified Multimodal Generation}},
      author={Bo Li and Yida Yin and Wenhao Chai and Xingyu Fu and Zhuang Liu},
      year={2026},
      eprint={2601.22155},
      archivePrefix={arXiv},
      primaryClass={cs.CV},
      url={https://arxiv.org/abs/2601.22155},
}

@misc{latentumm,
      title={{LatentUMM: Dual Latent Alignment for Unified Multimodal Models}},
      author={Yinyi Luo and Wenwen Wang and Hayes Bai and Marios Savvides and Jindong Wang},
      year={2026},
      eprint={2605.17766},
      archivePrefix={arXiv},
      url={https://arxiv.org/abs/2605.17766},
}

@inproceedings{unipath,
  title={{UniPath: Adaptive Coordination of Understanding and Generation for Unified Multimodal Reasoning}},
  author={Hayes Bai and Yinyi Luo and Wenwen Wang and Qingsong Wen and Jindong Wang},
  booktitle={Advances in Neural Information Processing Systems (NeurIPS)},
  year={2026}
}

@inproceedings{torchumm,
  title={{TorchUMM: A Unified Multimodal Model Codebase for Evaluation, Analysis, and Post-training}},
  author={Yinyi Luo and Wenwen Wang and Hayes Bai and Hongyu Zhu and Hao Chen and Pan He and Marios Savvides and Sharon Li and Jindong Wang},
  booktitle={Advances in Neural Information Processing Systems (NeurIPS)},
  year={2026}
}

@misc{chameleon,
      title={{Chameleon: Mixed-Modal Early-Fusion Foundation Models}},
      author={{Chameleon Team}},
      year={2024},
      eprint={2405.09818},
      archivePrefix={arXiv},
      url={https://arxiv.org/abs/2405.09818},
}

@inproceedings{showo,
  author = {Xie, Jinheng and Mao, Weijia and Bai, Zechen and Zhang, David Junhao and Wang, Weihao and Lin, Kevin Qinghong and Gu, Yuchao and Chen, Zhijie and Yang, Zhenheng and Shou, Mike Zheng},
  title = {{Show-o: One Single Transformer to Unify Multimodal Understanding and Generation}},
  booktitle = {International Conference on Learning Representations},
  pages = {28240--28264},
  year = {2025},
  url = {https://proceedings.iclr.cc/paper_files/paper/2025/hash/45f0d179ef7e10eb7366550cd4e574ae-Abstract-Conference.html}
}

@article{emu3,
  author = {Wang, Xinlong and Cui, Yufeng and Wang, Jinsheng and Zhang, Fan and Wang, Yueze and Zhang, Xiaosong and Luo, Zhengxiong and Sun, Quan and Li, Zhen and Wang, Yuqi and Yu, Qiying and Zhao, Yingli and Ao, Yulong and Min, Xuebin and Men, Chunlei and Wu, Boya and Zhao, Bo and Zhang, Bowen and Wang, Liangdong and Liu, Guang and He, Zheqi and Yang, Xi and Liu, Jingjing and Lin, Yonghua and Wang, Zhongyuan and Huang, Tiejun},
  title = {Multimodal learning with next-token prediction for large multimodal models},
  journal = {Nature},
  volume = {650},
  number = {8101},
  pages = {327--333},
  year = {2026},
  doi = {10.1038/s41586-025-10041-x},
  url = {https://www.nature.com/articles/s41586-025-10041-x}
}

@misc{transfusion,
      title={{Transfusion: Predict the Next Token and Diffuse Images with One Multi-Modal Model}},
      author={Chunting Zhou and Lili Yu and Arun Babu and Kushal Tirumala and Michihiro Yasunaga and Leonid Shamis and Jacob Kahn and Xuezhe Ma and Luke Zettlemoyer and Omer Levy},
      year={2024},
      eprint={2408.11039},
      archivePrefix={arXiv},
      url={https://arxiv.org/abs/2408.11039},
}

@inproceedings{negclip,
      title={{When and why Vision-Language Models behave like Bags-of-Words, and what to do about it?}},
      author={Mert Yuksekgonul and Federico Bianchi and Pratyusha Kalluri and Dan Jurafsky and James Zou},
      booktitle={International Conference on Learning Representations},
      year={2023},
      url={https://openreview.net/forum?id=KRLUvxh8uaX},
}

@inproceedings{winoground,
      title={{Winoground: Probing Vision and Language Models for Visio-Linguistic Compositionality}},
      author={Tristan Thrush and Ryan Jiang and Max Bartolo and Amanpreet Singh and Adina Williams and Douwe Kiela and Candace Ross},
      booktitle={Proceedings of the IEEE/CVF Conference on Computer Vision and Pattern Recognition},
      year={2022},
      pages={5238--5248},
      url={https://arxiv.org/abs/2204.03162},
}

@inproceedings{sugarcrepe,
      title={{SugarCrepe: Fixing Hackable Benchmarks for Vision-Language Compositionality}},
      author={Cheng-Yu Hsieh and Jieyu Zhang and Zixian Ma and Aniruddha Kembhavi and Ranjay Krishna},
      booktitle={Thirty-Seventh Conference on Neural Information Processing Systems Datasets and Benchmarks Track},
      year={2023},
      url={https://arxiv.org/abs/2306.14610},
}

@article{zou2026advancing,
  title = {Advancing Aesthetic Image Generation via Composition Transfer},
  author = {Zou, Kai and Zhao, Zhiwei and Liu, Bin and Yu, Nenghai},
  journal = {International Journal of Computer Vision},
  volume = {134},
  pages = {252},
  year = {2026},
  doi = {10.1007/s11263-026-02862-8},
  url = {https://doi.org/10.1007/s11263-026-02862-8},
}

@misc{qwen25vl,
      title={{Qwen2.5-VL Technical Report}},
      author={Shuai Bai and Keqin Chen and Xuejing Liu and Jialin Wang and Wenbin Ge and Sibo Song and Kai Dang and Peng Wang and Shijie Wang and Jun Tang and Humen Zhong and Yuanzhi Zhu and Mingkun Yang and Zhaohai Li and Jianqiang Wan and Pengfei Wang and Wei Ding and Zheren Fu and Yiheng Xu and Jiabo Ye and Xi Zhang and Tianbao Xie and Zesen Cheng and Hang Zhang and Zhibo Yang and Haiyang Xu and Junyang Lin},
      year={2025},
      eprint={2502.13923},
      archivePrefix={arXiv},
      primaryClass={cs.CV},
      url={https://arxiv.org/abs/2502.13923},
}

@misc{qwen3,
      title={{Qwen3 Technical Report}},
      author={An Yang and Anfeng Li and Baosong Yang and Beichen Zhang and Binyuan Hui and Bo Zheng and Bowen Yu and Chang Gao and Chengen Huang and Chenxu Lv and Chujie Zheng and Dayiheng Liu and Fan Zhou and Fei Huang and Feng Hu and Hao Ge and Haoran Wei and Huan Lin and Jialong Tang and Jian Yang and Jianhong Tu and Jianwei Zhang and Jianxin Yang and Jiaxi Yang and Jing Zhou and Jingren Zhou and Junyang Lin and Kai Dang and Keqin Bao and Kexin Yang and Le Yu and Lianghao Deng and Mei Li and Mingfeng Xue and Mingze Li and Pei Zhang and Peng Wang and Qin Zhu and Rui Men and Ruize Gao and Shixuan Liu and Shuang Luo and Tianhao Li and Tianyi Tang and Wenbiao Yin and Xingzhang Ren and Xinyu Wang and Xinyu Zhang and Xuancheng Ren and Yang Fan and Yang Su and Yichang Zhang and Yinger Zhang and Yu Wan and Yuqiong Liu and Zekun Wang and Zeyu Cui and Zhenru Zhang and Zhipeng Zhou and Zihan Qiu},
      year={2025},
      eprint={2505.09388},
      archivePrefix={arXiv},
      primaryClass={cs.CL},
      url={https://arxiv.org/abs/2505.09388},
}

@misc{mmvet,
      title={{MM-Vet: Evaluating Large Multimodal Models for Integrated Capabilities}},
      author={Weihao Yu and Zhengyuan Yang and Linjie Li and Jianfeng Wang and Kevin Lin and Zicheng Liu and Xinchao Wang and Lijuan Wang},
      year={2023},
      eprint={2308.02490},
      archivePrefix={arXiv},
      url={https://arxiv.org/abs/2308.02490},
}

@inproceedings{irg,
  author = {Huang, Wenxuan and Chen, Shuang and Xie, Zheyong and Cao, Shaosheng and Tang, Shixiang and Shen, Yufan and Yin, Qingyu and Hu, Wenbo and Wang, Xiaoman and Tang, Yuntian and Qiao, Junbo and Guo, Hangyu and Hu, Yao and Yin, Zhenfei and Torr, Philip and Cheng, Yu and Ouyang, Wanli and Lin, Shaohui},
  title = {{Interleaving Reasoning for Better Text-to-Image Generation}},
  booktitle = {International Conference on Learning Representations},
  pages = {106153--106182},
  year = {2026},
  url = {https://proceedings.iclr.cc/paper_files/paper/2026/hash/ad48f017e6c3d474caf511208e600459-Abstract-Conference.html}
}

@inproceedings{unison,
  title = {{Unison}: Benchmarking Unified Multimodal Models via Synergistic Understanding and Generation},
  author = {Liu, Jinyu and Shuai, Xincheng and Ding, Henghui and Jiang, Yu-Gang},
  booktitle = {International Conference on Machine Learning},
  year = {2026},
  url = {https://github.com/FudanCVL/Unison}
}

@inproceedings{janus,
  author = {Wu, Chengyue and Chen, Xiaokang and Wu, Zhiyu and Ma, Yiyang and Liu, Xingchao and Pan, Zizheng and Liu, Wen and Xie, Zhenda and Yu, Xingkai and Ruan, Chong and Luo, Ping},
  title = {{Janus}: Decoupling Visual Encoding for Unified Multimodal Understanding and Generation},
  year = {2025},
  booktitle = {Proceedings of the IEEE/CVF Conference on Computer Vision and Pattern Recognition (CVPR)},
  pages = {12966--12977},
  url = {https://openaccess.thecvf.com/content/CVPR2025/html/Wu_Janus_Decoupling_Visual_Encoding_for_Unified_Multimodal_Understanding_and_Generation_CVPR_2025_paper.html}
}

@inproceedings{janusflow,
  author = {Ma, Yiyang and Liu, Xingchao and Chen, Xiaokang and Liu, Wen and Wu, Chengyue and Wu, Zhiyu and Pan, Zizheng and Xie, Zhenda and Zhang, Haowei and Yu, Xingkai and Zhao, Liang and Wang, Yisong and Liu, Jiaying and Ruan, Chong},
  title = {{JanusFlow}: Harmonizing Autoregression and Rectified Flow for Unified Multimodal Understanding and Generation},
  year = {2025},
  booktitle = {Proceedings of the IEEE/CVF Conference on Computer Vision and Pattern Recognition (CVPR)},
  pages = {7739--7751},
  url = {https://openaccess.thecvf.com/content/CVPR2025/html/Ma_JanusFlow_Harmonizing_Autoregression_and_Rectified_Flow_for_Unified_Multimodal_Understanding_CVPR_2025_paper.html}
}

@inproceedings{omnigen2,
  author = {Wu, Chenyuan and Wang, Jiahao and Zheng, Pengfei and Yan, Ruiran and Xiao, Shitao and Luo, Xin and Wang, Yueze and Li, Wanli and Jiang, Xiyan and Liu, Yexin and Zhou, Junjie and Xia, Ziyi and Liu, Ze and Li, Chaofan and Deng, Haoge and Luo, Kun and Zhang, Bo and Zhang, Jiajun and Liu, Dong and Lian, Defu and Wang, Xinlong and Wang, Zhongyuan and Huang, Tiejun and Liu, Zheng},
  title = {{OmniGen2}: Towards Instruction-Aligned Multimodal Generation},
  year = {2026},
  booktitle = {Proceedings of the IEEE/CVF Conference on Computer Vision and Pattern Recognition (CVPR)},
  pages = {21964--21975},
  url = {https://openaccess.thecvf.com/content/CVPR2026/html/Wu_OmniGen2_Towards_Instruction-Aligned_Multimodal_Generation_CVPR_2026_paper.html}
}

@inproceedings{mmada,
  author = {Yang, Ling and Tian, Ye and Li, Bowen and Zhang, Xinchen and Shen, Ke and Tong, Yunhai and Wang, Mengdi},
  title = {{MMaDA}: Multimodal Large Diffusion Language Models},
  year = {2025},
  booktitle = {Advances in Neural Information Processing Systems},
  volume = {38},
  pages = {138867--138907},
  doi = {10.52202/085713-4636},
  url = {https://proceedings.neurips.cc/paper_files/paper/2025/hash/caa934a507a952698d54efb24845fc4b-Abstract-Conference.html}
}

@article{ovisu1,
  author = {Wang, Guo-Hua and Zhao, Shanshan and Zhang, Xinjie and Cao, Liangfu and Zhan, Pengxin and Duan, Lunhao and Lu, Shiyin and Fu, Minghao and Chen, Xiaohao and Zhao, Jianshan and Li, Yang and Chen, Qing-Guo},
  title = {{Ovis-U1} Technical Report},
  year = {2025},
  journal = {arXiv preprint arXiv:2506.23044},
  url = {https://arxiv.org/abs/2506.23044}
}

@article{emu35,
  author = {Cui, Yufeng and Chen, Honghao and Deng, Haoge and Huang, Xu and Li, Xinghang and Liu, Jirong and Liu, Yang and Luo, Zhuoyan and Wang, Jinsheng and Wang, Wenxuan and Wang, Yueze and Wang, Chengyuan and Zhang, Fan and Zhao, Yingli and Pan, Ting and Li, Xianduo and Hao, Zecheng and Ma, Wenxuan and Chen, Zhuo and Ao, Yulong and Huang, Tiejun and Wang, Zhongyuan and Wang, Xinlong},
  title = {{Emu3.5}: Native Multimodal Models are World Learners},
  year = {2025},
  journal = {arXiv preprint arXiv:2510.26583},
  url = {https://arxiv.org/abs/2510.26583}
}

@article{luo2026flexumm,
  title = {{FlexUMM}: Flexible Representation Learning for Unified Multimodal Models},
  author = {Luo, Yinyi and Bai, Haoyue and Ge, Yi and Vales, Nilton Cesar Rojas and Savvides, Marios and Wang, Jindong},
  journal = {arXiv preprint arXiv:2610.22682},
  year = {2026},
  url = {https://arxiv.org/abs/2610.22682}
}

@article{luo2026beyond,
  title = {Beyond Structural Unification: A Perspective and Roadmap on Unified Multimodal Models},
  author = {Luo, Yinyi and Bai, Hayes and Jin, Yiqiao and Hua, Wenyue and Kumar, Srijan and Savvides, Marios and Wen, Qingsong and Wang, Jindong},
  journal = {arXiv preprint arXiv:2610.35906},
  year = {2026},
  url = {https://arxiv.org/abs/2610.35906}
}
